\documentclass[prd, aps, nofootinbib, preprint, letterpaper]{revtex4-2}

\usepackage{mathtools}
\usepackage{color}
\usepackage{graphicx}
\usepackage{tabularx}
\definecolor{darkred}{RGB}{196,0,0}

\usepackage{amsmath, amssymb, amsfonts}

\usepackage[hidelinks]{hyperref}
\usepackage{ulem}
    
\newcommand{\be}{\begin{equation}}
\newcommand{\ee}{\end{equation}}
\newcommand{\ba}{\begin{eqnarray}}
\newcommand{\ea}{\end{eqnarray}}

\newcommand{\beq}{\begin{equation}}
\newcommand{\eeq}{\end{equation}}
\newcommand{\bqa}{\begin{eqnarray}}
\newcommand{\eqa}{\end{eqnarray}}
\newcommand{\la}{\label}

\begin{document}

\title {Influence of a nontrivial Polyakov loop on the quarkonium states}

\author{Wenqiang Liu,$^{1,2}$ Lihua Dong,$^{3}$ and Yun Guo$^{1,2,*}$}

\affiliation{$^1$ Department of Physics, Guangxi Normal University,\\ Guilin, 541004, China \\
$^2$ Guangxi Key Laboratory of Nuclear Physics and Technology, \\Guilin, 541004, China \\
$^3$ School of Physics and Astronomy, Sun Yat-Sen University, \\Zhuhai, 519082, China\\}

\renewcommand{\thefootnote}{\fnsymbol{footnote}}
\footnotetext[1]{Contact author: yunguo@mailbox.gxnu.edu.cn}
\renewcommand{\thefootnote}{\arabic{footnote}}

\begin{abstract}
Using a hybrid approach based on a phenomenological heavy-quark potential model and the background field effective theory, we assess the influence of a nontrivial Polyakov loop on the in-medium properties of the heavy quarkonium states. Without resorting to any temperature-dependent parameter, the lattice simulations on the complex heavy-quark potential are well reproduced with the potential model in which the screening masses are determined by the background field effective theory. Because of the reduced screening strength near the deconfining temperature, a nontrivial Polyakov loop leads to a dramatic increase in the binding energies, together with a moderate decrease in the decay widths. In general, a more tightly bounded quarkonium state can be expected, although the increase in the dissociation temperatures is significant only for relatively large-size bounded states. Our results indicate that the background field modification on the binding and decay of the quarkonium states may have a notable impact on the density evolution of quarkonia during the expansion of the fireball, and thus on the final observed quarkonium spectra.

\end{abstract}
\maketitle
\newpage

\section{Introduction}

Because of the nonperturbative nature, thermodynamics near the deconfining phase transition in $SU(N)$ gauge theories poses a big challenge to physicists. The perturbation theory and its improved version, known as the hard-thermal-loop perturbation theory (HTLpt) work well in the high temperature limit, but fail to reproduce the lattice simulations on the thermodynamic quantities at temperatures close to the deconfining temperature $T_d$ \cite{Andersen:1999fw,Andersen:2002ey,Haque:2014rua}. In a semiquark-gluon plasma (semi-QGP), which refers to a partially deconfined phase located in a temperature region from $T_d$ to $\sim 3T_d$ \cite{Hidaka:2008dr}, in addition to lattice QCD simulations as a reliable way to explore nonperturbative physics, various effective theories have also been developed for the same purpose. 

On the other hand, heavy quarkonium dissociation was proposed a long time ago as a sensitive probe to study the hot medium created in ultrarelativistic heavy-ion experiments \cite{matsui1986j}. Quarkonium physics is most relevant in the nonperturbative semi-QGP because the most bounded quarkonium $\Upsilon(1S)$ is unlikely to survive above $\sim 3T_d$. Therefore, utilizing an effective theory to study the quarkonium physics appears feasible. The background field effective theory (BFET) \cite{Hidaka:2020vna} turns out to be an ideal candidate for the following reasons. The thermodynamics of semi-QGP is well described within the effective theory \cite{Dumitru:2010mj,Dumitru:2012fw,Guo:2014zra}, and thus the resulting screening behavior is improved as compared to HTLpt. This is particularly important because the screening strength plays a crucial role in determining the binding and decay of the heavy bound states. Furthermore, the BFET also predicts a nontrivial temperature dependence of the Polyakov loop, showing a discontinuity of the order parameter at the phase transition point for $SU(3)$ as observed in lattice simulations. Therefore, it is possible to investigate the impact of the deconfining phase transition on the physical properties of quarkonia.

The success of the potential nonrelativistic QCD justifies the description of heavy quarkonia in terms of a static potential \cite{Brambilla:1999xf,Brambilla:2004jw} and a quantum-mechanical treatment based on the Schr\"odinger equation becomes applicable. In this work, we adopt a hybrid approach based on a phenomenological heavy-quark potential model and the BFET to analyze the in-medium properties of the quarkonium states. We will focus on the influence of a nontrivial Polyakov loop on the complex heavy-quark potential, and thus on the binding energies and decay widths of charmonia and bottomonia in a semi-QGP. As the only temperature-dependent parameter in the phenomenological model, the screening mass is determined nonperturbatively within the effective theory. The validity of such a hybrid approach is verified by comparing the model predictions on the complex heavy-quark potential with the corresponding lattice data. The lattice simulation reported in \cite{Burnier:2016mxc} show
a screened real part of the static potential which is
supported by the traditional understanding from the theoretical point of view that color screening produced by the light quarks and gluons weakens the interaction between the quark-antiquark pair. However, more recent lattice data suggest an unscreened real part of the static potential \cite{Bazavov:2023dci,Larsen:2024wgw}, and a similar conclusion is also found through the application of a deep neural network \cite{Shi:2021qri}. The inconsistency among the different lattice data is found to be due to different hypotheses for the shapes of the spectral function \cite{Bala:2021fkm}. At present, we still need to wait for a conclusive result from lattice studies. In the current work, we will use the simulation data from \cite{Burnier:2016mxc} when making a comparison with our model predictions. 

The rest of the paper is organized as follows. In Sec.~\ref{model}, we introduce the improved Karsch-Mehr-Satz (KMS) potential model and the background field effective theory, which jointly determine the complex heavy-quark potential without using any temperature-dependent parameter. We also show the good agreement obtained by comparing model predictions with the lattice simulations. In Sec.~\ref{properties}, in-medium properties of the heavy-quark bound states under the given potential model are studied. We first consider bound states with extremely large quark masses, and some analytical results are derived with the quantum-mechanical perturbation theory. By solving the Schr\"odinger equation, we also obtain the binding energies and decay widths for several low-lying charmonium and bottomonium bound states, and discuss the influence of a nontrivial Polyakov loop on these physical quantities. We list some important conclusions drawn from this work in Sec.~\ref{con}.

\section{The complex heavy-quark potential in the semi-QGP}\label{model}

Because of the large quark masses, quarkonia can be studied by using the Schr\"odinger equation in which a proper nonrelativistic potential model is very essential in determining the physical properties of the bound states. 
At finite temperatures, many phenomenological models have been proposed. The well-known Karsch-Mehr-Satz (KMS) potential model \cite{Karsch:1987pv} and its improved versions take a form of the screened Cornell potential and have been widely used to study the in-medium properties of quarkonia. For example, an improved KMS potential model as proposed in \cite{Guo:2018vwy} reads\footnote{We first discuss the real part of the static potential and come to the imaginary part later.}
\be
\label{reV}
\mathrm{Re}\,V(\hat{r}) = - \alpha_s \bigg(m_D+\frac{e^{-\hat{r}}}{r}\bigg)+
\frac{2 \sigma}{m_D}\left[1-\exp\left( -\hat{r} \right)\right]-\frac{\sigma}{m_D}\hat{r} \exp\left( -\hat{r} \right)\, ,
\ee
where $r$ denotes the separation distance between the quark and antiquark, and $\hat {r}\equiv r m_D$ with $m_D$ being the Debye screening mass. In addition, the QCD coupling constant is given by $\alpha_s=g^2 C_F/(4\pi)$ and $\sigma$ is the string tension. The static potential can be defined by Fourier transforming the temporal component of the (retarded) gluon propagator $D_R(P)$ in the static limit
\be
\label{Vdef}
\mathrm{Re}\,V(\hat{r}) = -g^2 C_F \int \frac{d^3 \mathbf{p}}{(2\pi)^3} \left( e^{i {\mathbf{p}} \cdot {\mathbf{r}}}-1 \right) D_R(\omega=0,\mathbf{p})\, .
\ee
Accordingly, the potential model in Eq.~(\ref{reV}) indicates that the gluon propagator can be formally written as $D_R \equiv D_{R,c}+D_{R,s}$ with
\ba
\label{Vpre}
D_{R,c}(\omega=0,\mathbf{p})&=& \frac{1}{p^2+m_D^2} \, ,\nonumber \\ D_{R,s}(\omega=0,\mathbf{p})&=& \frac{m_G^2}{(p^2+m_D^2)^2}+\frac{4 m_G^2 m_D^2}{(p^2+m_D^2)^3}\, .
\ea
In the above equation, $m_G^2=2\sigma/\alpha_s$ is a dimensional constant related to the gluon condensate in vacuum \cite{Chetyrkin:1998yr}. The Coulomb part $D_{R,c}$ in the propagator can be derived from the HTLpt by considering gluon self-energy insertion into the bare propagator. The string part $D_{R,s}$ appearing as a phenomenological contribution dominates at large distances.

\begin{figure}[htbp]
\centering
\includegraphics[width=0.32\textwidth]{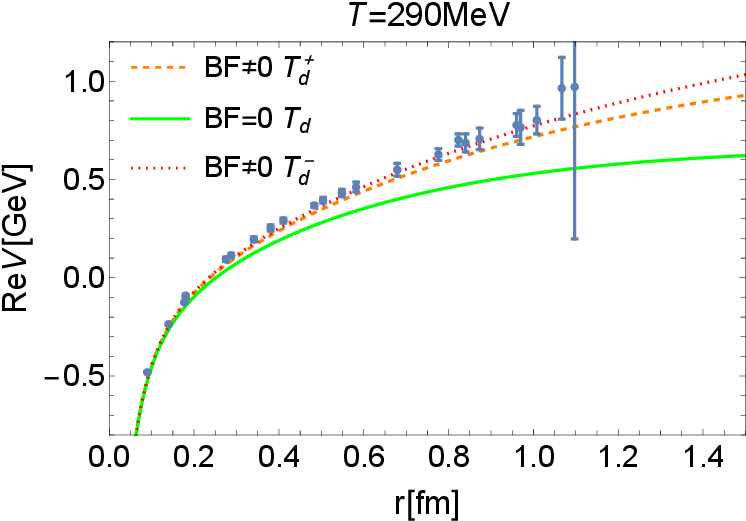}
\includegraphics[width=0.32\textwidth]{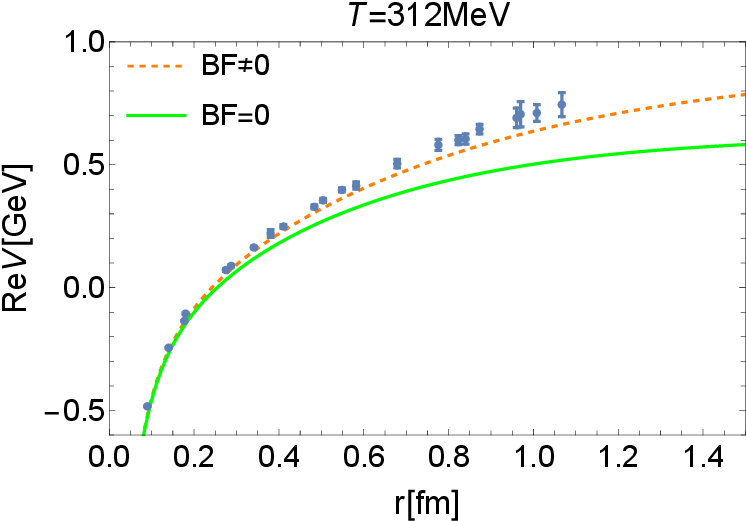}
\includegraphics[width=0.32\textwidth]{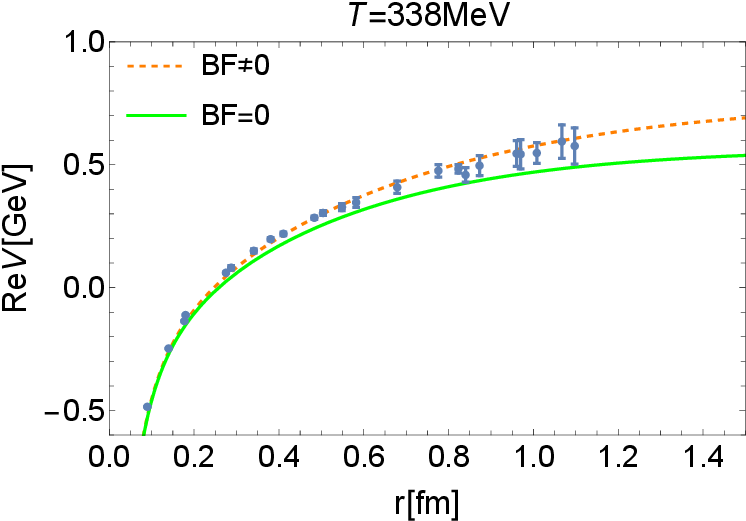}
\includegraphics[width=0.32\textwidth]{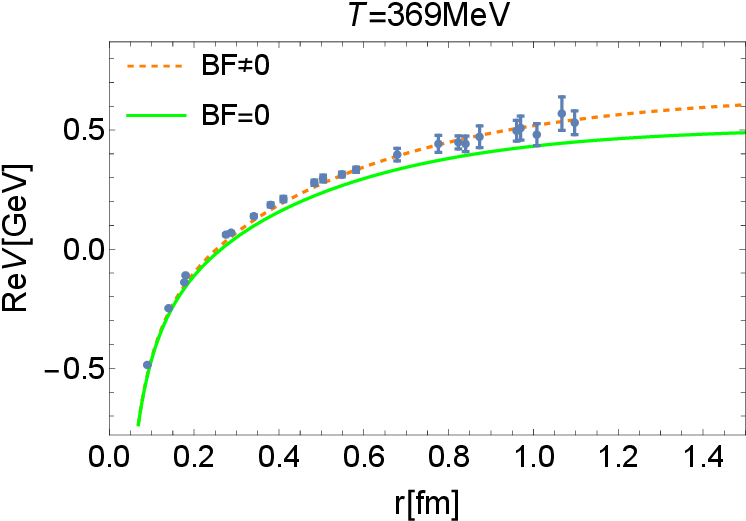}
\includegraphics[width=0.32\textwidth]{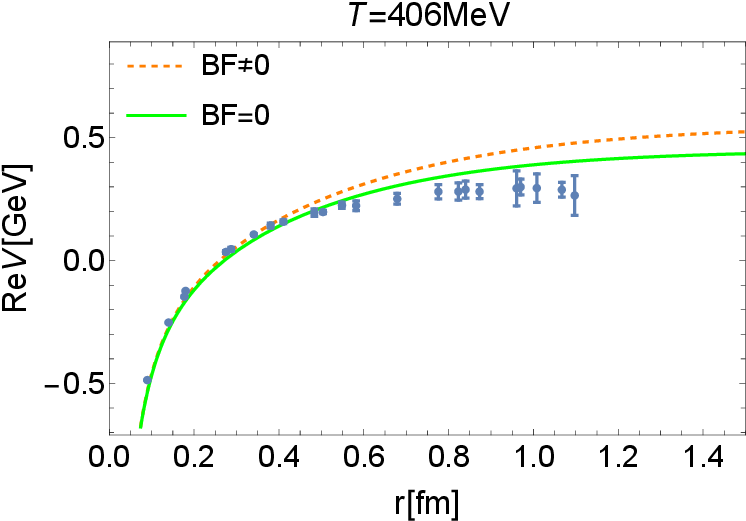}
\vspace*{0.15cm}
\caption{Comparisons of $\mathrm{Re}\,V$ between the lattice data from \cite{Burnier:2016mxc} and the improved KMS potential models at different temperatures. The model predictions based on the HTLpt and the BFET are denoted by the solid and dashed curves, respectively. The discontinuity of the potential at $T_d$ in the BFET is also shown in the first plot.}
\label{fig1}
\end{figure}

It can be checked that when the medium effect is removed, 
i.e., $m_D \rightarrow 0$, the improved KMS potential model is reduced to vacuum Cornell potential as required. For simplicity, the coupling constant and the string tension are assumed to be temperature independent which can be fixed by fitting the lattice potential at zero temperature.\footnote{The corresponding values are found to be $\alpha_s \approx 0.272$ and $\sigma\approx 0.215\,{\rm GeV}^2$.} Consequently, medium modifications on the Cornell potential are entirely encoded in the screening mass. On the other hand, as a key assumption in the KMS potential model, the very same screening scale as appears in the Coulomb term also shows up in the nonperturbative string contribution. Therefore, such a screening mass should be understood as a nonperturbative quantity for consistency. Although an infinite series of the perturbative Feynman diagrams are involved within the HTL resummation framework, it is still not a full nonperturbative treatment, especially for the physics at temperatures not far above $T_d$. Recall that the results of the thermodynamics from HTLpt cannot reproduce the corresponding lattice simulations in semi-QGP; it is therefore not surprising to see that the improved KMS potential model in Eq.~(\ref{reV}) gives an unsatisfactory description of the in-medium heavy-quark potential when the nonperturbative screening mass is determined based on HTLpt, which at leading order reads $m_D=g T$ for $SU(3)$ gauge theory.\footnote{We only consider $SU(3)$ pure gauge theory in this work and set $N_f=0$ to drop the contributions from dynamical quarks.} According to Fig.~\ref{fig1}, for temperatures close to $T_d$, the model predictions obviously undershoot the lattice data taken from \cite{Burnier:2016mxc}, although the agreements are gradually improved as the temperature increases. In these plots, the leading order screening mass is evaluated with the two-loop running coupling $g$ which is given by
\be
g^{-2}(T)=2\beta_0\ln\left(2\pi T/\Lambda_{\overline{MS}}\right)+\frac{\beta_1}{\beta_0}
\ln\left[2\ln\left(2\pi T/\Lambda_{\overline{MS}}\right)\right]\, ,
\label{2loop}
\ee
with $\beta_0=11/(16\pi^2)$ and $\beta_1=102/(256\pi^4)$. In numerical evaluations, we choose $\Lambda_{\overline{MS}}\approx 350\,{\rm MeV}$.

It is worth noting that if we take the Debye mass $m_D$ as a free parameter, Eq.~(\ref{reV}) can fit the lattice data very well, which in turn gives the value of the screening mass at a specific temperature \cite{Guo:2018vwy}. This fact justifies that, given the correct behaviors of the screening mass, such a screened form of the Cornell potential is able to provide a good reproduction of the lattice potential. Without resorting to the lattice data, we will try to make use of the screening mass obtained in the BFET to assess the possible improvement on the agreement between the lattice data and model predictions.

According to lattice QCD, the values of the Polyakov loop are nonzero but less than unity in the semi-QGP. To model such a partially deconfined phase, the BFET self-consistently introduces a nonzero classical background field $A_0^{\rm cl}$ for the gauge field. It is diagonal in the color space and can be written as $(A_0^{\rm cl})_{ab}\equiv \delta_{ab} {\cal Q}^a/g $. The background field ${\cal Q}$ is expected to have a proper temperature dependence, leading to a Polyakov loop that exhibits the desired behavior during the phase transition. However, this cannot be realized in the perturbation framework because the system is always in the completely deconfined phase according to the equation of motion that ${\cal Q}$ should obey. To drive the system to confinement, a nonperturbative contribution $\sim  m^2 T^2$ must be added to the perturbative effective potential $\sim T^4$. There are two ways to generate this contribution. One is to add a mass term $m^2$ in the inverse bare gluon propagator and expand the resulting effective potential by requiring $m\ll T$ \cite{Meisinger:2001cq}. The other is to consider the contribution from two-dimensional ghosts embedded isotropically in four dimensions where $m$ acts as an upper limit of the transverse momentum of the embedded ghost field \cite{Hidaka:2020vna}. In any case, $m$ is not a free parameter that can be fixed by requiring the deconfining phase transition to occur at $T_d$.

By analogy with the calculation in the HTLpt, we can rederive the resummed gluon propagator $D_{R,c}$ as given in Eq.~(\ref{Vpre}) within the BFET by inserting the ${\cal Q}$-dependent gluon self-energy into the bare propagator. The static limit of the retarded gluon propagator $ D_{R,c}^{aa,bb}(P)$ for the diagonal gluons in $SU(3)$ is given by \cite{Guo:2020jvc}
\be\la{diapro}
\sum_{ab}{\cal P}^{aa,bb}  D_{R,c}^{aa,bb}(\omega=0, {\bf p})=\frac{1}{p^2+\big({\cal M}^{(1)}_{D}\big)^2}+\frac{1}{p^2+\big({\cal M}^{(2)}_{D}\big)^2}\, .
\ee
In the above equation, we used the double line basis\footnote{Because of the overcompleteness of the generators in the double line basis, the individual component of the diagonal propagator $D_{R,c}^{aa,bb}$ cannot be uniquely determined, instead, only the sum as given in Eq.~(\ref{diapro}) has an unambiguous expression.} where the gluons are denoted by a pair of color indices $ab$ with $a,b=1,2,3$ and the projection operator is defined as ${\cal P}^{ab,cd}=\delta^{ad}\delta^{bc}-\delta^{ab}\delta^{cd}/3$. More details about the double line basis can be found in \cite{Hidaka:2009hs,Guo:2018scp}. The explicit forms of the ${\cal Q}$-modified screening masses associated with the two diagonal gluons are given by
\be\la{diamass}
{\cal M}^{(1)}_{D}= m_D\sqrt{1+\beta+ 2 s^2/3-2s}\,,\quad\quad{\rm and}\quad\quad {\cal M}^{(2)}_{D}= m_D \sqrt{1+\beta+ 2 s^2-10s/3}\,.
\ee
Here, $\beta=3 m^2/(4\pi^2 T^2)$ and $m^2=40\pi^2 T_d^2/81$. Throughout this work, we use $T_d=290\, {\rm MeV}$ as indicated by the lattice data of the heavy-quark potential \cite{Burnier:2016mxc}. The background field is a traceless matrix parametrized as ${\cal Q}=2 \pi T (s/3,0,-s/3)$ and its $T$ dependence can be obtained according to the equation of motion in the effective theory, $s=(3-\sqrt{9-24\beta})/4$. In fact, Eq.~(\ref{diamass}) shows the explicit ${\cal Q}$ modification on the screening mass $m_D$ in HTLpt. Furthermore, the propagator for off-diagonal gluons reads
\be\la{offpro}
D_{R,c}^{ab,ba}(\omega=0,{\bf p})=\frac{1}{p^2+\big({\cal M}^{(ab)}_{D}\big)^2}\, ,
\ee
with $a\neq b$. The corresponding ${\cal Q}$-modified screening masses associated with the six off-diagonal gluons in $SU(3)$ are given by
\be
{\cal M}_D^{(23)}={\cal M}_D^{(32)}={\cal M}_D^{(12)}={\cal M}_D^{(21)}=m_D\sqrt{1+\beta+7s^2/9-5s/3}\, ,\nonumber
\ee
\be\label{offmass}
{\cal M}_D^{(13)} = {\cal M}_D^{(31)}=m_D\sqrt{1+\beta+10s^2/9-2s}\,.
\ee

The screening mass in the HTLpt is diagonal in color space, namely, the $N^2-1$ gluons in $SU(N)$ acquire the same $m_D$. However, a nontrivial color structure appears in the presence of a background field, and the gluons become distinguishable by their associated screening masses. The screening behavior determined by HTLpt is reliable only when the temperature is far above $T_d$. On the other hand, we can naturally expect that the ${\cal Q}$-modified screening masses have an improved nonperturbative property because the effective theory can correctly describe the $T$-dependent Polyakov loop as well as the rapid change of the thermodynamics in the semi-QGP.

Adopting the key assumption used in the KMS potential model, i.e., the same screening scale appears in both Coulomb and string term, we can get the string part of the gluon propagator as follows:
\be
\label{newd}
\sum_{abcd}{\cal P}^{ab,cd}  D_{R,s}^{ab,cd}(\omega=0, {\bf p})= \sum_{i=1}^8 \frac{m_G^2}{\Big(p^2+\big({\cal M}^{[i]}_{D}\big)^2\Big)^2}+\frac{4 m_G^2 \big({\cal M}^{[i])}_{D}\big)^2}{\Big(p^2+\big({\cal M}^{[i]}_{D}\big)^2\Big)^3}\, .
\ee
In the above equation, the eight screening masses in $SU(3)$ as given in Eqs.~(\ref{diamass}) and (\ref{offmass}) are collectively denoted by ${\cal M}^{[i]}_{D}$ for simplicity. By rewriting Eq.~(\ref{Vpre}) in terms of the double line basis\footnote{We use $V$ and ${\tilde V}$ to denote the static potential with and without the background field modification, respectively. This also applies to other physical quantities, such as the binding energy $E$ and ${\tilde E}$, the decay width $\Gamma$ and ${\tilde {\Gamma}}$, etc.}
\be\la{potdef}
\mathrm{Re}\,{\tilde V} (\hat{r})=-\frac{g^2}{2N}\sum_{{\rm colors}} \int \frac{ d^3{\bf p}}{(2\pi)^3}(e^{i {\bf p} \cdot {\bf r}}-1){\cal P}^{ab,cd}D_{R}^{ab,cd}(\omega= 0,{\bf p})\,,
\ee
we find that the ${\cal Q}$ modification on the improved KMS potential model is very simple which only amounts to a replacement of the screening mass $m_D$ with the ${\cal Q}$-modified ones. Therefore, we arrive at
\be
\label{newmodel}
\mathrm{Re}\,{\tilde V}({{\hat r}}) = \frac{1}{8}\ {\sum^8_{i=1}}\bigg[- \alpha_s \Big({{\cal M}_D^{[i]}}+\frac{e^{-{\hat r}_{[i]}}}{r}\Big)+
\frac{2 \sigma}{{\cal M}_D^{[i]}}\big(1-\exp\left( -{\hat r}_{[i]} \right)\big)-\frac{\sigma}{{{\cal M}_D^{[i]}}}{{\hat r}_{[i]}} \exp\left( -{{\hat r}_{[i]}} \right)\bigg]\, ,
\ee
with ${\hat r}_{[i]}=r{{\cal M}_D^{[i]}} $.

To demonstrate how the ${\cal Q}$ modification plays a role in improving the model predictions on static potential, we make a comparison between the lattice data and the corresponding results from Eqs.~(\ref{reV}) and (\ref{newmodel}). As shown in Fig.~\ref{fig1}, without relying on any temperature dependent parameter,\footnote{Recall that the two parameters $\alpha_s$ and $\sigma$ are temperature independent and fixed by the lattice potential in vacuum.} a very good agreement with the data can be achieved based on Eq.~(\ref{newmodel}). Compared to the predictions from Eq.~(\ref{reV}), the improvement becomes significant as the temperature approaches $T_d$. This is because the influence of a nonzero background field is most accentuated in a narrow temperature region above $T_d$ where the quarkonium studies also become most relevant. Furthermore, at large separation distances, a sharp increase in the asymptotic values of the potential near the deconfining temperature can be found in the ${\cal Q}$-modified model which is very essential to determine the binding energy of the bound states. On the other hand, the background field vanishes very quickly with increasing temperature and has a negligible impact at high temperatures where HTLpt is sufficient to describe the screening behavior of the medium. However, at a relatively high temperature, for example $T=406\,{\rm MeV}$, a visible deviation from the data at large distances shows up which indicates that instead of considering the ${\cal Q}$ modification, contributions to the screening mass beyond the leading-order approximation in the HTLpt could be more important for a better agreement. Finally, it is worth pointing out that the effective theory predicts a discontinuity of the Polyakov loop at the deconfining temperature which reflects the nature of a first order phase transition in $SU(3)$. Therefore, the ${\cal Q}$-modified screening masses are not uniquely determined at this temperature and their values depend on how the temperature approaches $T_d$, from below $T\rightarrow T_d^-$ or from above $T\rightarrow T_d^+$. Consequently, at any specific distance, the static potential in Eq.~(\ref{newmodel}) is double valued when $T=T_d$, see the first plot in this figure. This phenomenon results in an impact on the in-medium properties of the quarkonia which will be further discussed later.

\begin{figure}[htbp]
\centering
\includegraphics[width=0.5\textwidth]{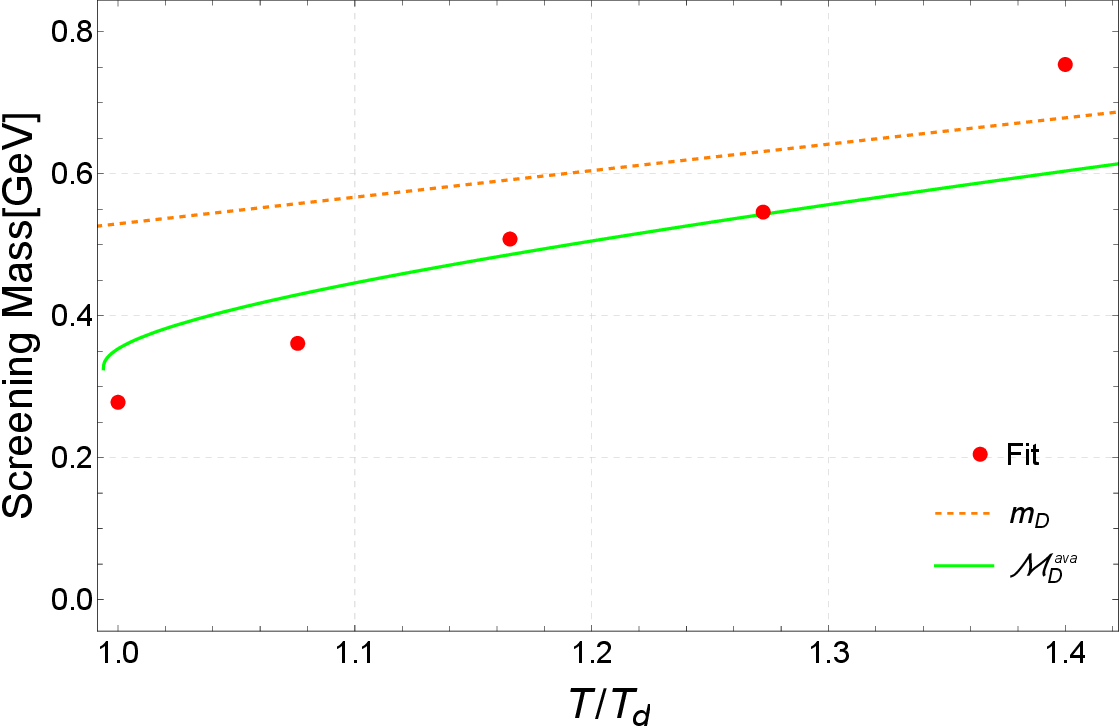}
\vspace*{0.1cm}
\caption{Comparison of the screening mass determined in different ways.}
\label{fig2}
\end{figure}

As the only $T$-dependent parameter in the improved KMS potential model, the screening mass is crucial to determine the behaviors of the in-medium static potential. Therefore, we also compare the screening masses from HTLpt ($m_D$) and BFET (${{\cal M}_D^{\rm ava}}$) with the corresponding results obtained by fitting the real part of the lattice potential based on the same phenomenological model. In order to make such a comparison, the screening mass ${{\cal M}_D^{\rm ava}}$ is actually the average value of the eight ${\cal Q}$-modified masses ${{\cal M}_D^{[i]}}$. As we can see in Fig.~\ref{fig2}, the averaged screening masses in BFET can reproduce the fitting results reasonably well, especially in a temperature region near $T_d$. This observation is consistent with the improvements demonstrated in Fig.~\ref{fig1}.

An important feature of the heavy-quark potential at finite temperature is that it develops an imaginary part, leading to a finite decay width of the quarkonium state in a hot medium. After replacing the retarded gluon propagator $D_R(P)$ in Eq.~(\ref{Vdef}) by the symmetric one $D_F(P)$, the imaginary part of the potential can be obtained through the following Fourier transform \cite{Guo:2018vwy} 
\be
\label{Videf}
\mathrm{Im}\,V(\hat{r}) = -g^2 C_F \int \frac{d^3 \mathbf{p}}{(2\pi)^3} \left( e^{i {\mathbf{p}} \cdot {\mathbf{r}}}-1 \right) \frac{1}{2}D_F(\omega=0,\mathbf{p})\, ,
\ee
with $D_F \equiv D_{F,c}+D_{F,s}$ and
\ba
\label{Vpim}
D_{F,c}(\omega=0,\mathbf{p})&=& -\frac{2\pi T m_D^2}{p(p^2+m_D^2)^2} \, ,\nonumber \\ D_{F,s}(\omega=0,\mathbf{p})&=& -12\pi T \frac{m_G^2 m_D^2}{p}\Big[\frac{2 m_D^2}{(p^2+m_D^2)^4}-\frac{1}{(p^2+m_D^2)^3}\Big]\, .
\ea

As long as the exact form of the retarded/advanced gluon propagator is specified, the symmetric propagator can be obtained through the following relation
\be\label{kms}
D_F(P)=\big(1+2n(\omega)\big)\,{\rm sgn}(\omega)\big(D_R(P)-D_A(P)\big)\, ,
\ee
where $n(\omega)$ denotes the Bose-Einstein distribution and ${\rm sgn}(\omega)$ is the sign function. This is actually the Dyson-Schwinger equation in thermal equilibrium, where the retarded, advanced, and symmetric gluon self-energies satisfy the Kubo-Martin-Schwinger condition. To derive the static limit of the string contribution to the symmetric propagator, according to Eq.~(\ref{kms}), we need to assume a proper $\omega$ dependence in the nonperturbative string part of the retarded/advanced propagator. This is because in the small $\omega$ limit, $n(\omega)\sim 1/\omega$, and thus the leading order term in the small $\omega$ expansion of $D_R(P)-D_A(P)$ should be proportional to $\omega$. In the improved KMS model, the same $\omega$ dependence as appears in the perturbative $D_{R/A,c}(P)$ has also been used in $D_{R/A,s}(P)$, that is, the following replacement should be applied to the static propagator $D_{R/A,s}(\omega=0,\mathbf{p})$:
\be\label{rep}
m^2_D \rightarrow  m^2_D \Big(\frac{\omega}{2p}\ln\frac{\omega+p\pm i \epsilon}{\omega - p\pm i \epsilon}-1\Big)\,. 
\ee

The explicit form of the symmetric propagator as given in Eq.~(\ref{Vpim}) leads to the following imaginary part of static potential:
\be
\label{imVold}
\mathrm{Im}\,V(\hat{r}) = - \alpha_s T \phi_2(\hat{r}) +\frac{8\sigma T}{m_D^2} \phi_3(\hat{r})- \frac{24\sigma T}{m_D^2} \phi_4(\hat{r})\,.
\ee
In the above equation, the function $\phi_n(\hat{r})$ is defined as
\begin{equation}
\phi_n(\hat{r}) = 2 \int_0^\infty dz \frac{z}{(z^2 +1)^n} \left[1- \frac{\mathrm{sin}(z \hat{r})}{z \hat{r}}\right]\, .
\end{equation}

Generalizing the above calculation to the semi-QGP turns out to be straightforward because the ${\cal Q}$-modified retarded/advanced propagator is already known and the $\omega$ dependence can be introduced through a similar replacement as given in Eq.~(\ref{rep}) for each individual screening mass. However, since the propagator is not diagonal in color space when using the double line basis, we rewrite Eq.~(\ref{kms}) with the explicit color indices as
\be\label{kmsnew}
\sum_{\rm colors}{\cal P}^{ab,cd}D^{ab,cd}_F(P)=\big(1+2n(\omega)\big)\,{\rm sgn}(\omega)\sum_{\rm colors}{\cal P}^{ab,cd}\big(D^{ab,cd}_R(P)-D^{ab,cd}_A(P)\big)\, .
\ee
Accordingly, the imaginary part of the heavy-quark potential can be obtained by a similar Fourier transform as given in Eq.~(\ref{potdef}) by changing $D_{R}^{ab,cd}$ into $D_{F}^{ab,cd}/2$. Notice that a background field affects the distributions of the thermal partons with typical momentum $\sim T$. However, for soft partons with typical momentum $\sim gT$, we neglect the possible corrections to their distribution in the above equation. This makes the resulting symmetric propagator $D_{F}^{ab,cd}(P)$ analogous to that as given in Eq.~(\ref{Vpim}) when taking the static limit. In addition, to get the above relation, we assume that the Kubo-Martin-Schwinger condition also holds in the BFET although it has only been verified in a perturbative way by considering a nonzero background field \cite{Wang:2022dcw}.

\begin{figure}[htbp]
\centering
\includegraphics[width=0.32\textwidth]{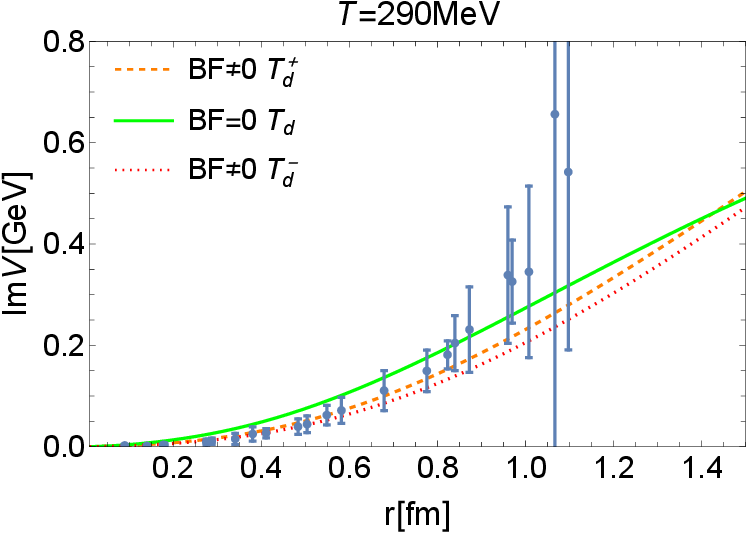}
\includegraphics[width=0.32\textwidth]{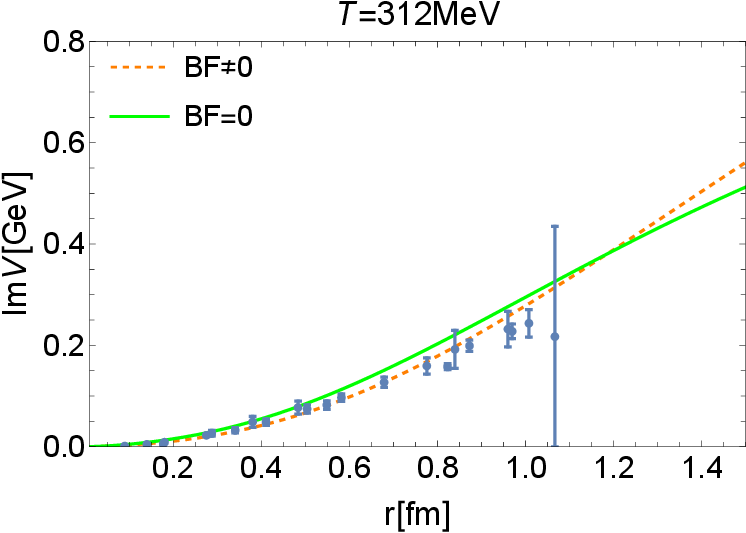}
\includegraphics[width=0.32\textwidth]{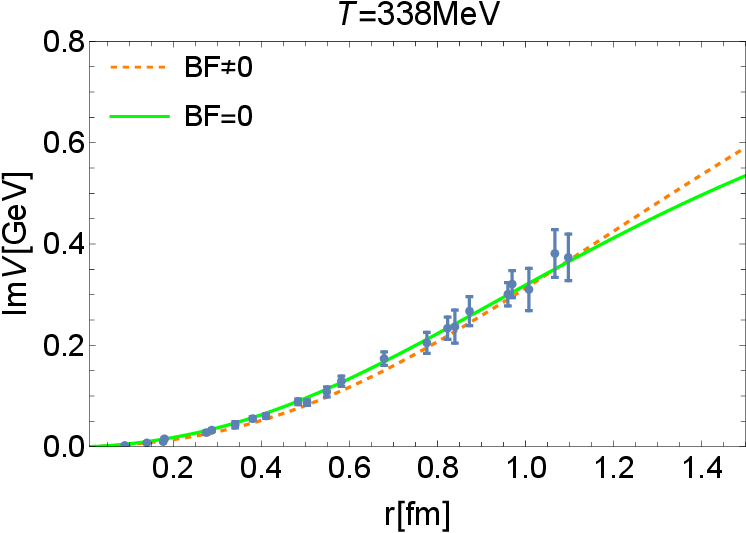}
\includegraphics[width=0.32\textwidth]{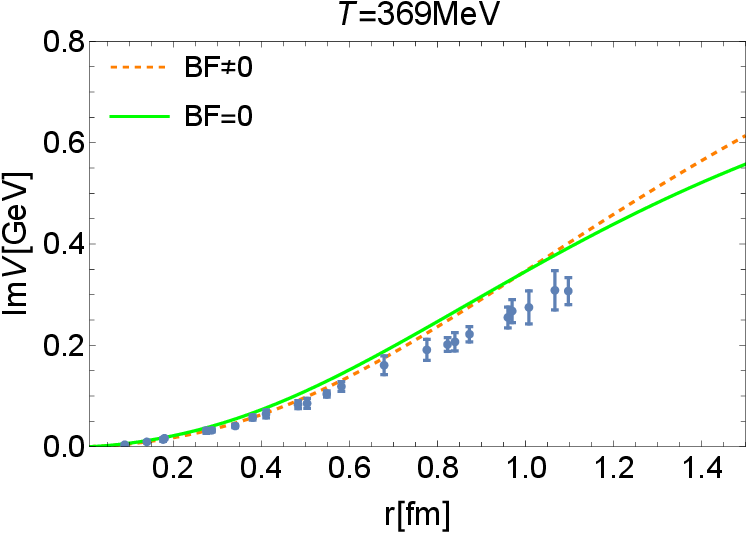}
\includegraphics[width=0.32\textwidth]{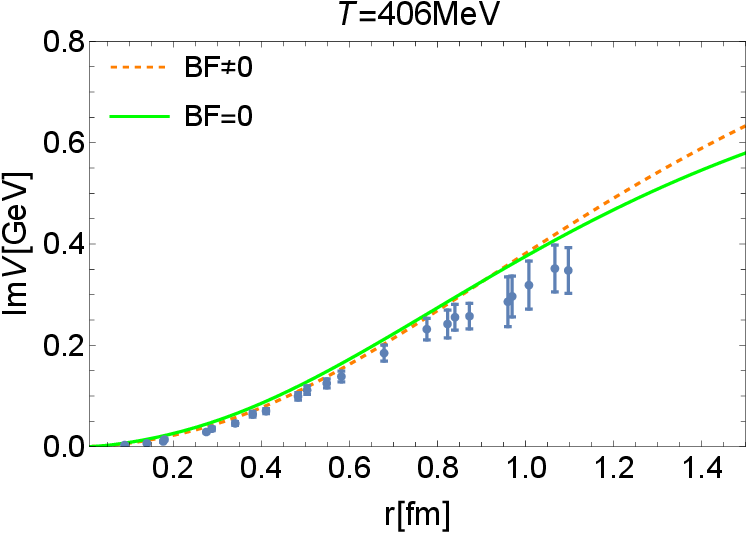}
\vspace*{0.1cm}
\caption{Comparisons of $\mathrm{Im}\,V$ between the lattice data from \cite{Burnier:2016mxc} and the improved KMS potential models at different temperatures. The model predictions based on the HTLpt and the BFET are denoted by the solid and dashed curves, respectively. The discontinuity of the potential at $T_d$ in the BFET is also shown in the first plot.}
\label{fig3}
\end{figure}

Given the above discussions, we can get the imaginary part of the static potential in the presence of a nonzero background field,
\be
\label{imv}
{\rm{Im}}\,{\tilde V}(\hat{r}) =\frac{1}{8}\ {\sum^8_{i=1}}\bigg(- \alpha_s T \phi_2({{\hat r}_{[i]}}) +\frac{8\sigma T}{\big({{\cal M}_D^{[i]}}\big)^2} \phi_3({{\hat r}_{[i]}})- \frac{24\sigma T}{\big({{\cal M}_D^{[i]}}\big)^2} \phi_4({{\hat r}_{[i]}})\bigg)\,.
\ee
Similar as the real part of the potential, we only need a replacement of $m_D$ with the ${\cal Q}$-modified ${\cal M}_D^{[i]}$ when switching from the HTLpt to the BFET. Given the improved KMS potential model, such a replacement can be naturally expected because the change in the medium properties in semi-QGP should be reflected in the modification of the screening mass. By comparing the results from Eqs.~(\ref{imVold}) and~(\ref{imv}) with the lattice data, Fig.~\ref{fig3} shows the corresponding improvements in the imaginary part of the static potential. It is found that in general, better agreement with the data can be obtained after taking into account the ${\cal Q}$ modification. The influence of the background field becomes most significant near the deconfining temperature, and a satisfactory reproduction of the lattice data can be realized in a wide temperature region, especially for small separation distances between the heavy quark and antiquark. Despite the large uncertainties in the lattice simulation as the distance increases, we see a discrepancy between the data and model predictions at large distances. However, quarkonium states that can survive above $T_d$ are bounded in a size that cannot be very large according to previous studies \cite{Dumitru:2009ni}, therefore, this discrepancy appears to be irrelevant for our studies.

Previously, the BFET has been used to study the collisional energy loss of a heavy quark in semi-QGP \cite{Du:2024riq}. In comparison to the results obtained from HTLpt, the suppression of the energy loss indicates a strong nonperturbative effect. The above comparisons with the first principle calculations indirectly justify the use of the effective theory, especially the nonperturbative behaviors of the screening masses near the deconfining temperature.
 
Finally, we mention that in addition to the KMS potential model, there are some other phenomenological models on the market \cite{Thakur:2013nia,Burnier:2015nsa}. In these models, the medium effect is introduced through the dielectric function which is directly related to the resummed gluon propagator. As a result, the ${\cal Q}$ modification can be taken into account in these models through a similar replacement of the Debye mass. From the point of view of model construction, the model in \cite{Thakur:2013nia} appears simpler as it relies on fewer model assumptions, i.e., the potential model is obtained by correcting the Cornell potential through the complex dielectric function. For consistency, such a dielectric function should be determined nonperturbatively. However, neither HTLpt nor BFET can determine a reasonable dielectric function by which the model would reproduce the lattice data. 
In fact, we find that this potential model requires an even smaller screening mass than the ${\cal Q}$-modified ones to fit the data for the real part of the potential; however, a simultaneous fit to both the real and imaginary parts seems unlikely.

\section{In-medium Properties of quarkonia in the presence of a background field}\label{properties}

In the presence of a nontrivial Polyakov loop, we have shown that there exists a significant ${\cal Q}$ modification on the complex heavy-quark potential in the semi-QGP. It would be interesting to study the corresponding modifications on the in-medium properties of a quarkonium state, such as the binding energy, decay width, and dissociation temperature. To do so, we first consider a bound state with an extremely large quark mass so that the ${\cal Q}$ modification can be obtained through the quantum-mechanical perturbation theory. For realistic quark masses, we determine the in-medium properties of charmonia and bottomonia by numerically solving the Schr\"odinger equation based on the improved KMS potential model with and without a background field.

\subsection{Medium corrections to a Coulombic state}
\label{lm}

For an extremely large quark mass $M$, the quark and antiquark are tightly bounded with a Bohr radius $\sim 1/(\alpha_s M)$ much smaller than the screening length $\sim 1/m_D$. For ${\hat r}\ll 1$, the medium effect can be considered as a small perturbation to the vacuum Coulomb potential. Therefore, we can expand the real part of the potential as the following
\be\label{bf1r0}
{\rm{Re}}\, V ({\hat r} \ll 1)=-\frac{\alpha_s}{r}+
\Big(\frac{\sigma}{m_D}-\frac{\alpha_s m_D}{2}\Big){\hat r}+{\cal O}({\hat r}^2) \, .
\ee
The static potential at infinitely large distance approaches $-\alpha_s m_D+2 \sigma/m_D$, and thus the binding energy for the ground state $1S$ is given by\footnote{We actually consider the absolute values of the binding energies.} 
\ba\label{bf1r1e1}
E =\frac{\alpha_s^2 M}{4}-\alpha_s m_D+\frac{2 \sigma}{m_D}+
\frac{3(\alpha_sm_D^2-2 \sigma)}{2 \alpha_s M}\, ,
\ea
where the last term is the medium induced correction to the eigenenergy $-\alpha_s^2 M/4$ of a Coulombic state.

\begin{figure}[htbp]
\centering
\includegraphics[width=0.5\textwidth]{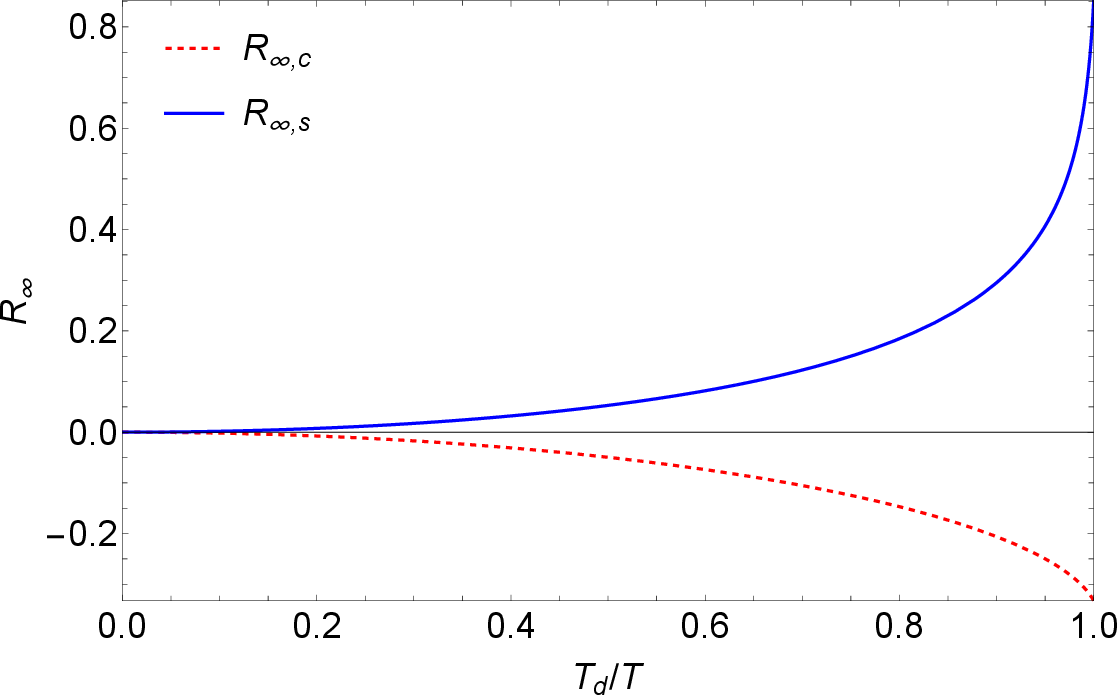}
\vspace*{0.1cm}
\caption{Temperature dependence of the ${\cal Q}$ modification on the (real part of the) static potential at infinity which is denoted by $R_\infty$. Results are shown for the Coulomb part $R_{\infty,c}$ and string part $R_{\infty,s}$, separately.}
\label{fig4}
\end{figure}

In the presence of a nonzero background field, the corresponding binding energy ${\tilde E}$ can be obtained from Eq.~(\ref{bf1r1e1}) by making the following change on those terms depend on $m_D$
\be\label{repla}
{\cal F}(m_D)\rightarrow \frac{1}{8} \sum_{i=1}^8  {\cal F}({\cal M}_D^{[i]}) \, .
\ee
The leading-order contribution to ${\tilde E}-E$ comes from the ${\cal Q}$ modification on the (real part of the) static potential at infinity which has no dependence on the quark mass and can be studied through the ratio $R_\infty \equiv ({\rm Re}\,{\tilde V}_\infty-{\rm Re}\, V_\infty)/ {\rm Re}\,V_\infty$. In Fig.~\ref{fig4}, we show the temperature dependence of $R_\infty$ for the Coulomb and string parts separately. Because of the weakened screening strength in semi-QGP, the increase in the potential at infinity indicates an increased binding energy. In particular, near the deconfining temperature, the background field has a dramatic effect on the string part of $R_\infty$ that can reach $\sim 85 \%$ as $T\rightarrow T_d$. Consequently, the change in the binding energy is approximately equal to $1.7 \sigma/m_D$. Although it is quantitatively small as compared to the eigenenergy in the large $M$ limit, such a change may play an important role in determining the binding energies for quarkonia with finite quark masses. 

On the other hand, the ${\cal Q}$ modification on the eigenenergy gives rise to the subleading contribution to ${\tilde E}-E$ which is suppressed by a factor of $\sim m_D/(\alpha_s M)$. At this order, the background field does not affect the string term in the potential; therefore, the modification leads to a decrease in the binding energy. In the high-temperature limit where the background field behaves as $\sim (T_d/T)^2$, we find a simple expression for the ${\cal Q}$-modified binding energy,
\be\label{zz2}
{\tilde E}\approx E+\frac{5}{27} \Big(\alpha_s m_D+\frac{2 \sigma}{m_D}\Big)\Big(\frac{T_d}{T}\Big)^2-\frac{5}{9}\frac{m_D^2}{M}\Big(\frac{T_d}{T}\Big)^2 \, .
\ee

The above analysis also applies to excited states and similar conclusions can be drawn. Furthermore, we point out that higher order correction to the Coulomb potential as given in Eq.~(\ref{bf1r0}) leads to a split of the binding energy for the first excited state. A nonzero background field will increase the energy gap between the $1P$ and $2S$ states by $14m_D^2/27 M$ as $T\rightarrow T_d$.

At finite temperature, the complex heavy-quark potential results in an imaginary part for the eigenenergies that directly relates to the decay width $\Gamma$ of the bound states. For extremely large $M$, we can also treat the imaginary part of the potential as a perturbation to the vacuum Coulomb potential. For small ${\hat r}$, the leading order term in ${\rm Im}\, V$ reads
\be\label{zy1}
{\rm Im}\, V ({\hat r}\ll 1) = \frac{\alpha_s T }{3} {\hat r}^2 \ln {\hat r}\, ,
\ee
which gives the following decay width for the ground state $1S$
\be\label{zy2}
\Gamma =\frac{4 m_D^2 T }{\alpha_s M^2}\ln \frac{\alpha_s M}{2 m_D} \, .
\ee
Notice that the contribution to $\Gamma$ from the string term is suppressed by $\sim m^2_D/(\alpha_s M)^2$ as compared to the above equation, and thus dropped in our calculation. The background field modification on the decay width can be obtained similarly by using Eq.~(\ref{repla}). In the limit $\alpha_s M\gg m_D$, the ${\cal Q}$ modification on the screening mass $m_D$ inside the logarithm can be neglected and the change of the decay width can be expressed by the following ratio 
\be\label{zy3}
R_\Gamma=\frac{{\tilde \Gamma}-\Gamma}{\Gamma}=\frac{1}{24}\sqrt{81-80 \Big(\frac{T_d}{T}\Big)^2}-\frac{5 }{27}\Big(\frac{T_d}{T}\Big)^2-\frac{3}{8} \, .
\ee
The above result shows that there is a reduced decay width in the presence of a nonzero background field. At the deconfining temperature, the influence is very large and $|R_\Gamma| \sim 50 \%$. Equation (\ref{zy3}) is also valid for excited states. We also find that although the difference of $\Gamma$ between the $2S$ and $1P$ states is reduced due to the background field, however, the ratio $(\Gamma_{2S}-\Gamma_{1P})/\Gamma_{2S}=2/7$ remains in a semi-QGP. 

The above discussions are applicable in the deconfined phase; therefore, the temperature approaches $T_d$ from above, i.e., ${T\rightarrow T_d^+}$. As already mentioned before, the discontinuity of the Polyakov loop at $T_d$ corresponds to a double-valued potential; it is thus important to study the behaviors of the heavy bound states as ${T\rightarrow T_d^-}$ where the background field $s=1$. We determine the jump in both the binding energy and the decay width when the deconfining phase transition occurs,
\be
\la{jump}
{\tilde E}_{T_d^-} \approx  {\tilde E}_{T_d^+}+1.06 \frac{\sigma}{m_D}\, ,\quad\quad{\rm and}\quad\quad
{\tilde \Gamma}_{T_d^-} \approx  0.77 {\tilde \Gamma}_{T_d^+}\, .
\ee
In the above equation, we only consider the leading order contribution to the decay width, while for the binding energy, the dominant contribution comes from the jump in ${\rm Re}\,V_\infty$ where a contribution $\sim 0.10 \alpha_s m_D$ from the Coulomb term is dropped because, for typical values of the screening mass, it is much smaller than the corresponding contribution from the string term. When transiting to the confined phase, there is a $\sim 20 \%$ decrease in the decay width; and the binding energy increases by $\sim 1.06\sigma/m_D$ which, in the large quark mass limit, is negligible compared to the magnitude of the binding energy itself $\sim \alpha_s^2 M$. However, a substantial effect on the binding of quarkonium states with finite quark masses can be expected.

Based on the above results, quarkonium states are more tightly bounded when the influence of a nonzero background field is taken into account. For very large quark masses, the decrease in the decay widths is power suppressed as compared to the increase in the binding energies. 
As a result, higher dissociation temperatures in a semi-QGP are mainly attributed to the increased binding energies. It must be noted that conclusions in this subsection are only for qualitative purposes. A quantitative assessment on the ${\cal Q}$ modification on the in-medium properties of bottomonia and charmonia requires numerical solving of the Schr\"odinger equation, which will be discussed in the next subsection.

\subsection {The properties of charmonia and bottomonia in the semi-QGP}\label{cbt2}

Using a parallelized solver called quantumFDTD \cite{Strickland:2009ft,Delgado:2020ozh}, the Schr\"odinger equation is numerically solved with the improved KMS potential models. For comparison, we consider potential models with and without a background field whose explicit forms can be found in Eqs.~(\ref{reV}) and (\ref{newmodel}) for the real part and in Eqs.~(\ref{imVold}) and (\ref{imv}) for the imaginary part. The eigen/binding energies and the decay widths for several low-lying quarkonium bound states are obtained at various temperatures located in semi-QGP. In the numerical evaluations, the lattice size of $L = N a \approx 2.56\,{\rm fm}$ with a lattice spacing $a = 0.050\,{\rm GeV}^{-1} \approx 0.01\, {\rm fm}$ is used for the ground state of bottomonia $\Upsilon(1S)$. For the first excited state of bottomonia $\chi_b(1P)$ and the ground state of charmonia $J/\Psi$, we choose $a=0.085\,{\rm GeV}^{-1}$ and $N = 256$, corresponding to a lattice size of $4.35\,{\rm fm}$. In addition, the quark masses are given by $M_c=1.3\,{\rm GeV}$ and $M_b=4.7\,{\rm GeV}$. 

\begin{figure}[htbp]
\centering
\includegraphics[width=0.31\textwidth]{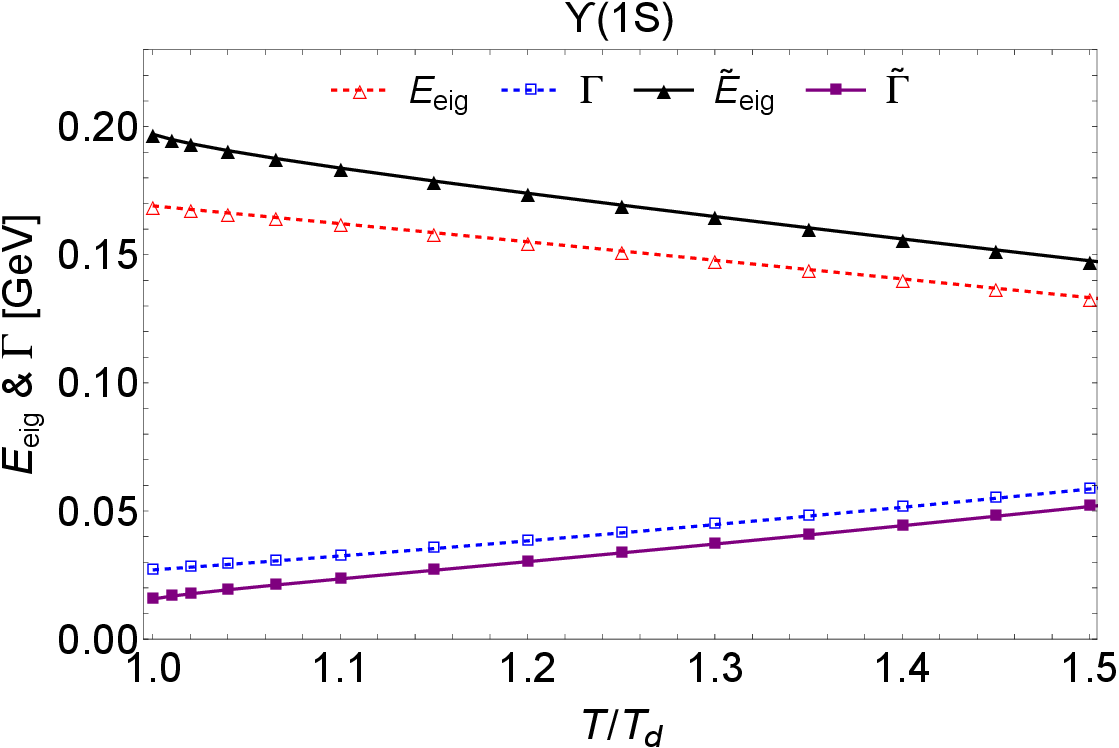}
\includegraphics[width=0.31\textwidth]{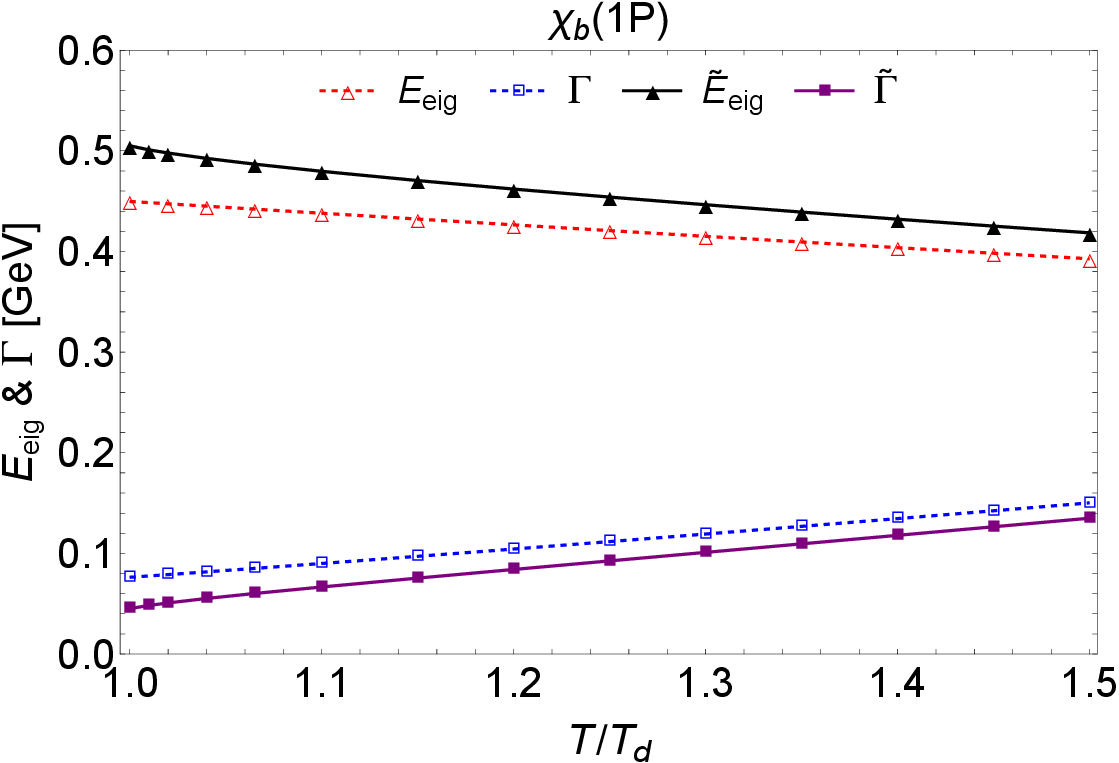}
\includegraphics[width=0.31\textwidth]{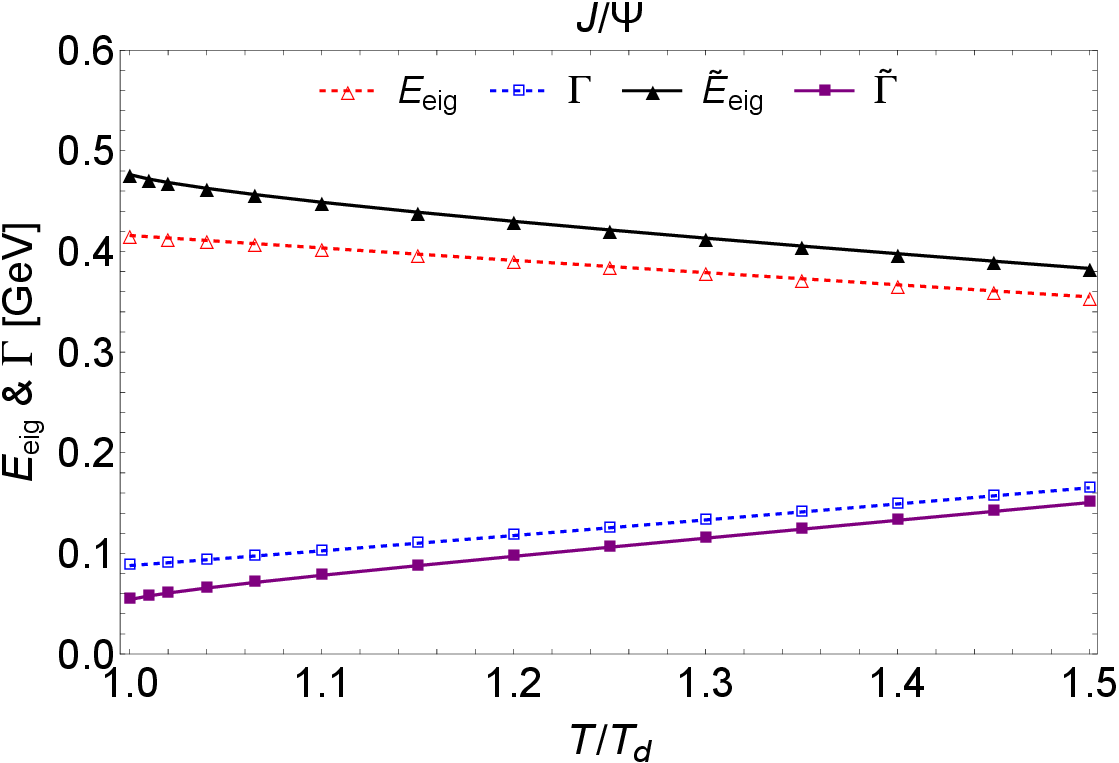}
\vspace*{0.1cm}
\caption{Temperature dependence of the eigenenergies and decay widths with and without the ${\cal Q}$ modifications for different quarkonium states. }
\label{fig5}
\end{figure}

\begin{figure}[htbp]
\centering
\includegraphics[width=0.32\textwidth]{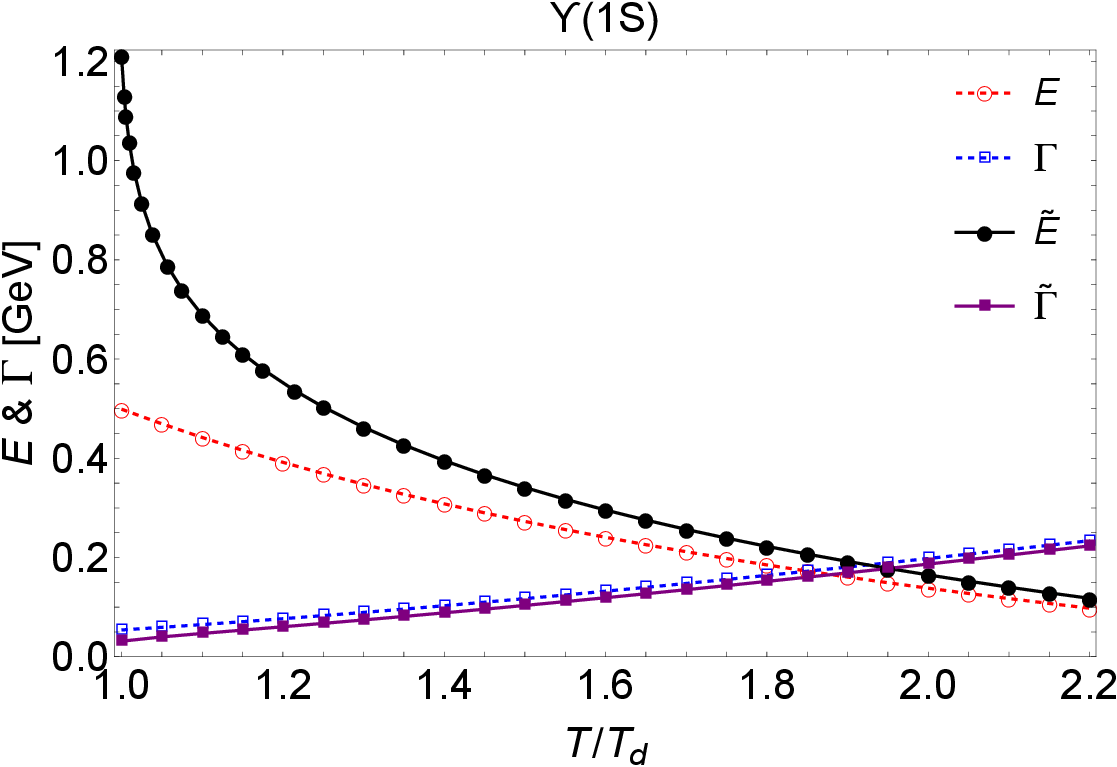}
\includegraphics[width=0.32\textwidth]{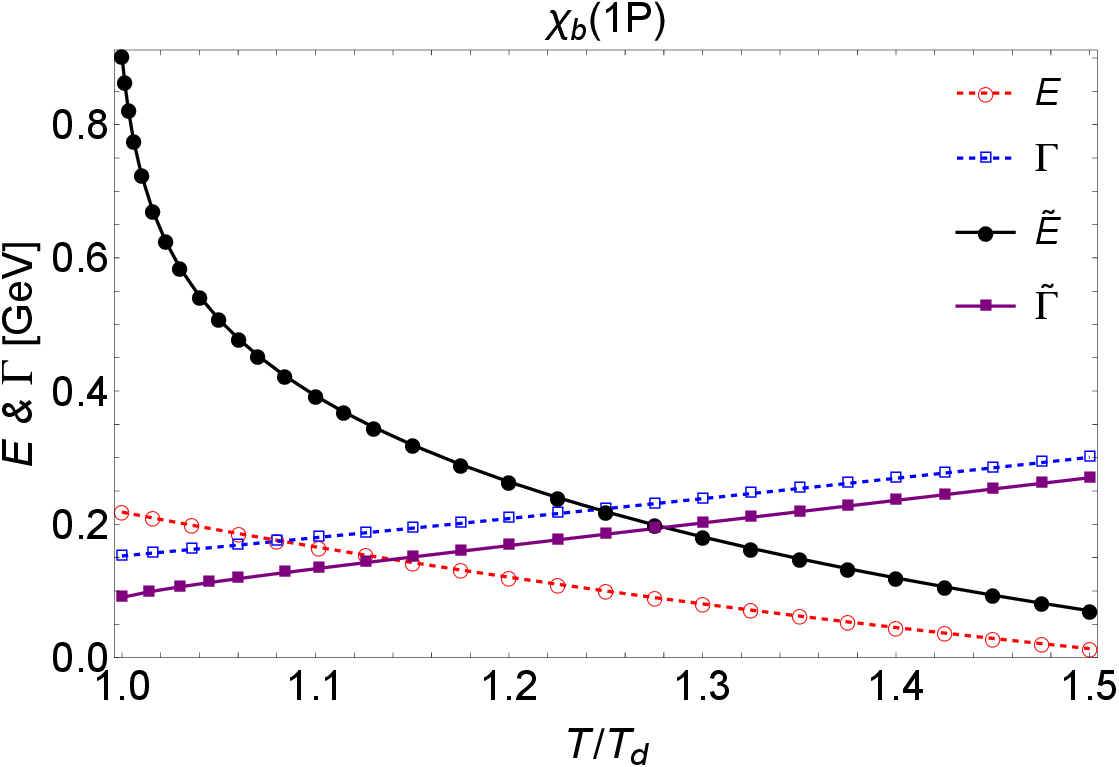}
\includegraphics[width=0.32\textwidth]{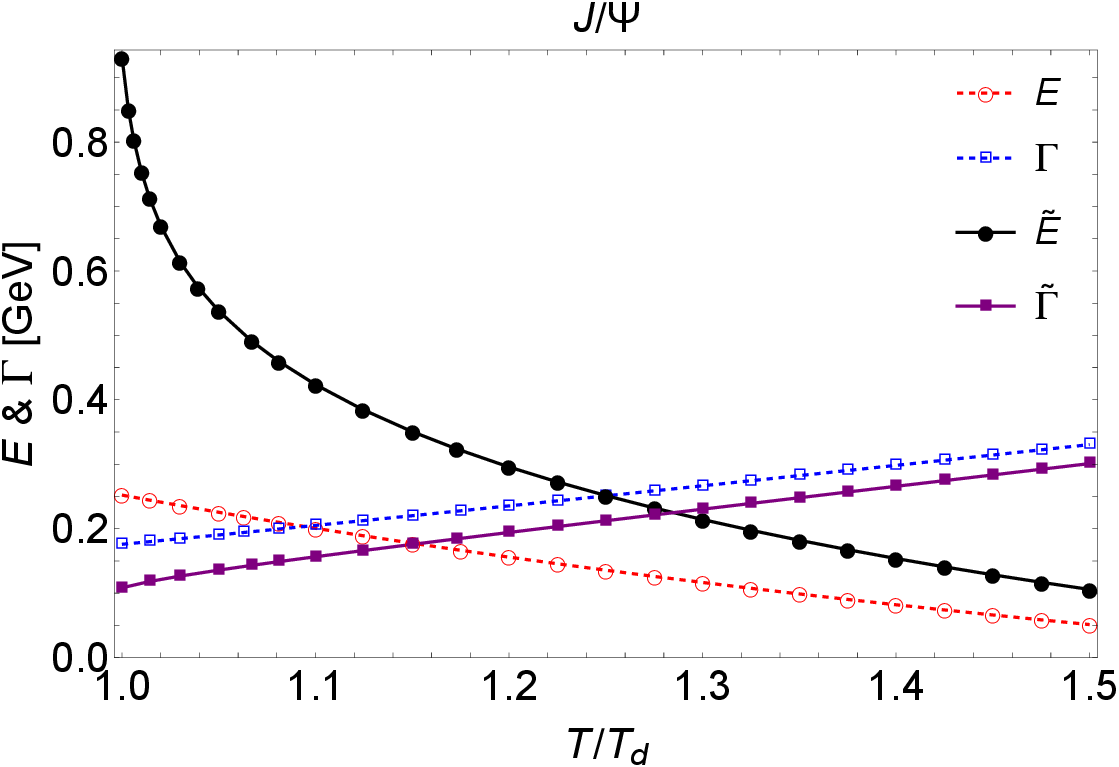}
\vspace*{0.1cm}
\caption{Temperature dependence of the binding energies and decay widths with and without the ${\cal Q}$ modifications for different quarkonium states. As compared to Fig.~\ref{fig5}, the values of the decay widths are doubled in this figure. 
}
\label{fig6}
\end{figure}

According to the results in Fig.~\ref{fig5}, there is a moderate decrease/increase appearing in the decay widths/eigenenergies after taking into account the background field. With a similar root-mean-square radius, the quantitative changes are almost the same for quarkonium states $\chi_b(1P)$ and $J/\Psi$, but reduced for the more bounded $\Upsilon(1S)$. On the other hand, the background field induces a remarkable modification on the binding energies which can be found in Fig.~\ref{fig6}. As the temperature approaches $T_d$, an increase $\sim 0.7\,{\rm GeV}$ in the binding energies is found for all the quarkonium states under consideration. Interestingly, this amount approximately equals the corresponding increase in the static potential at infinity. In contrast, the quantitative change in the eigenenergies is less than $\sim 10 \%$ of the amount increased in the binding energies as shown in Fig.~\ref{fig5}. Therefore, the ${\cal Q}$ modification on the binding energy is mainly caused by the large increase in ${\rm Re}\,V_\infty$ near the transition point.  Notice that in order to make the dissociation temperatures visible, we double the values of decay widths in Fig.~\ref{fig6}, so that the cross points of ${\tilde E}$ and ${\tilde \Gamma}$ or $E$ and $\Gamma$ determine the dissociation temperatures with or without the ${\cal Q}$ modification, respectively. It is clear that the nonzero background field also affects the dissociation temperatures which increase as a consequence of the increased $E$ together with the decreased $\Gamma$. For $\chi_b(1P)$ and $J/\Psi$, the dissociation temperature is roughly increased by $\sim 20 \%$. However, the corresponding influence becomes much weaker for $\Upsilon(1S)$ due to its higher dissociation temperature, at which the background field is too small to induce a considerable modification. 

\begin{figure}[htbp]
\centering
\includegraphics[width=0.32\textwidth]{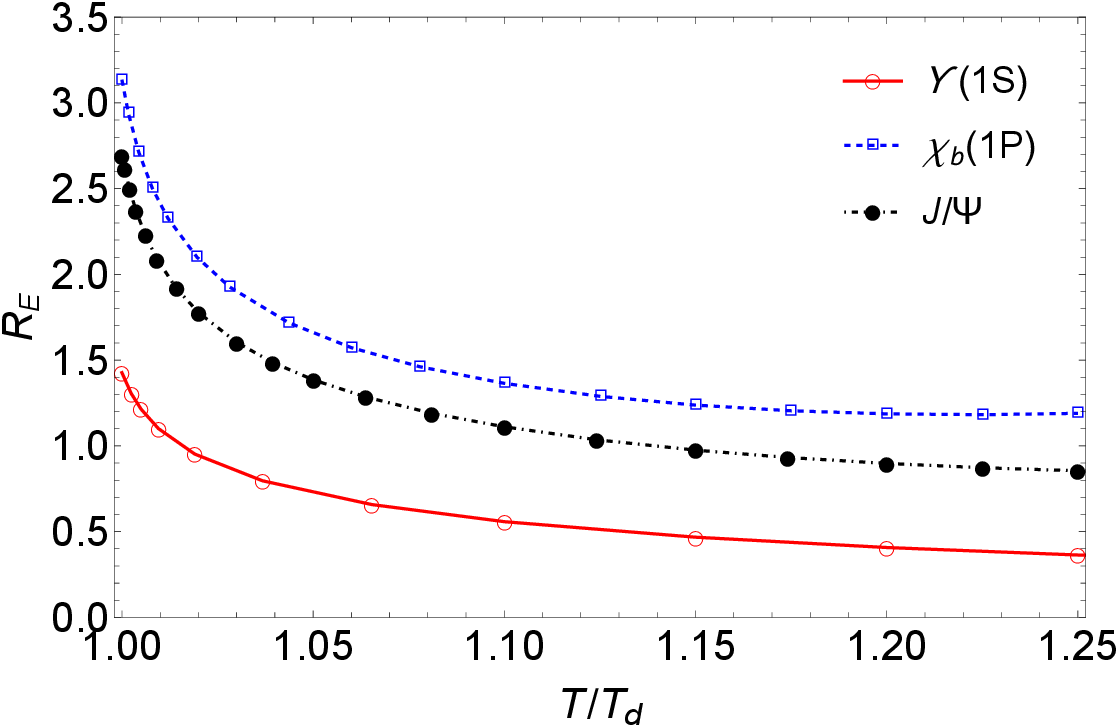}
\includegraphics[width=0.32\textwidth]{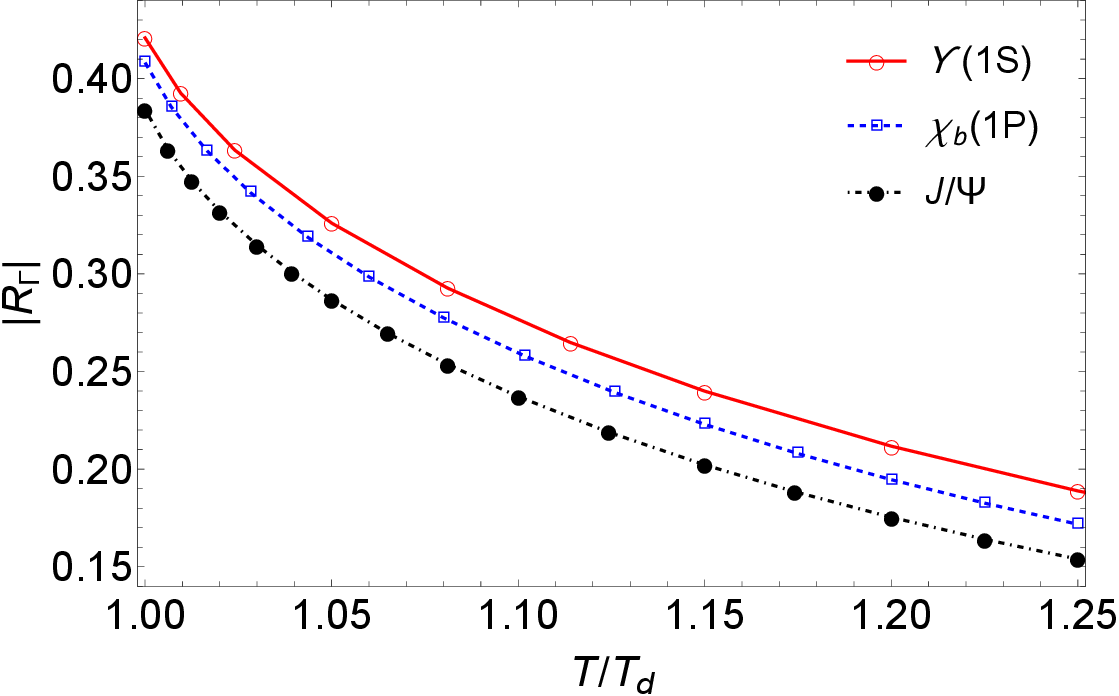}
\includegraphics[width=0.32\textwidth]{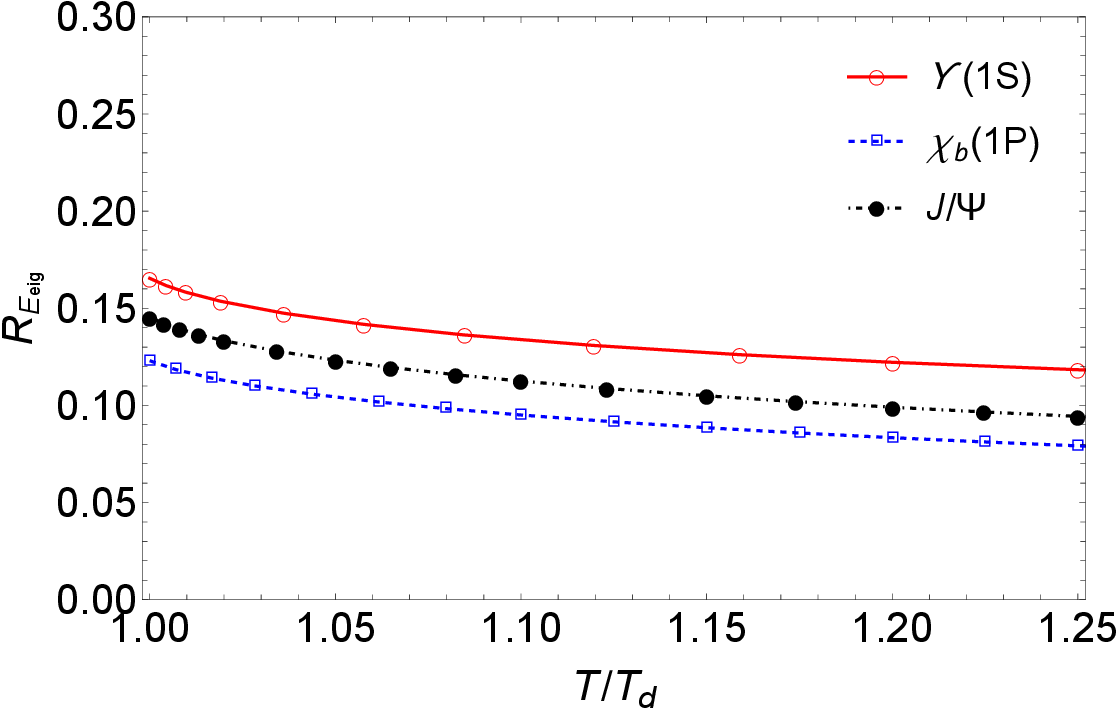}
\vspace*{0.1cm}
\caption{The relative corrections to the binding energies $R_E$, the decay widths $R_\Gamma$, and the eigenenergies $R_{E_{eig}}$ as a function of the scaled temperature for different quarkonium states. The definitions of the three ratios can be found in the text.}
\label{fig7}
\end{figure}

In the presence of a background field, the resulting change in ${\rm Re}\,V_\infty$ being dominant over the corresponding change in $E_{\rm eig}$ actually coincides with our finding in Sec.~\ref{lm} for the extremely heavy bound states. However, for finite quark masses, the relative correction to the binding energy defined as $R_E\equiv({\tilde E}-E)/E$ is no longer negligible. According to Fig.~\ref{fig7}, near the deconfining temperature, the ratio $R_E$ can reach $150 \%$ for $\Upsilon(1S)$, which is even larger for $\chi_b(1P)$ and $J/\Psi$. Therefore, the thermal density of the quarkonium states, being exponentially proportional to $E/T$, will be strongly enhanced by the background field. This enhancement only persists in a narrow temperature region because $R_E$ drops very quickly with increasing temperature, which is consistent with the very different slopes of ${\tilde E}$ and $E$ as observed in Fig.~\ref{fig6}. On the other hand, despite a moderate decrease in magnitude, the relative correction to the decay width increases to $\sim 40\%$ as approaching the deconfining temperature.\footnote{Recall that the absolute value of $R_\Gamma$ approaches $\sim 50\%$ as $T\rightarrow T_d$ for extremely heavy bound states.} Although not as significant as the relative correction to the binding energy, the ${\cal Q}$ modification may also play a role in determining the decay of the bound states and thus in describing the in-medium suppression of the final quarkonium spectra. We also mention that the relative correction to the eigenenergy $R_{E_{\rm eig}}\equiv ({\tilde E}_{\rm eig}-E_{\rm eig})/E_{\rm eig}$ shows a similar $T$-dependent behavior as $R_\Gamma$, however, the maximum of $R_{E_{\rm eig}}$ does not exceed $\sim 20 \%$ as found in Fig.~\ref{fig7}, indicating a weakened influence from the background field.

Finally, we also consider the discontinuities in the binding energies and decay widths at the deconfining temperature. The results are presented in Table \ref{dist}, which turn out to be consistent with the previous conclusions for the extremely heavy bound states as given in Eq.~(\ref{jump}). The jump in the binding energies and decay widths will further affect the thermal density of the quarkonium states, which could be quite substantial given the significant change, especially in the binding energies. For realistic QCD with dynamical quarks, the deconfining phase transition becomes a crossover. Although these discontinuities would disappear, one can still expect a rather different binding and decay behavior in semi-QGP, and thus an important modification on the density evolution of the quarkonium states towards the phase transition point, as compared to the case with a vanishing background field. 

\begin{table}[htpb]
\begin{center}
\setlength{\tabcolsep}{5mm}{
\begin{tabular}{   |c|  c|  c| }
\hline
(${\tilde E}$, ${\tilde \Gamma}$)      & $T\rightarrow T_d^-$  &  $T\rightarrow T_d^+$     \\ \hline
$  \Upsilon(1S)$ & $(1.647,0.013)$  & $ (1.211,0.016)             $     \\  \hline
$\chi_{b}(1P)$  & $(1.333,0.037)$  & $(0.903,0.045)         $     \\  \hline
$J/\Psi$  & $(1.360,0.044)$  & $(0.932,0.054)       $     \\   \hline	

\end{tabular}
}
\end{center}
\caption{The binding energies and decay widths at the phase transition point for different quarkonium states. We consider the temperature approaching $T_d$ from below $T\rightarrow T_d^-$ and from above $T\rightarrow T_d^+$. All results are given in units of ${\rm GeV}$.}
\label{dist}
\end{table}

\section{Conclusions}
\label{con}

As the order parameter of the deconfining phase transition, nontrivial values of the Polyakov loop near the deconfining temperature indicate a partial deconfinement of the strongly interacting medium and could be described by a classical background field $A_0^{\rm cl}$ for the gauge potential. In this work, we investigated the in-medium properties of heavy quarkonium states in such a partially deconfined plasma. 

We adopted the improved KMS potential model, which contains the Debye screening mass as the only $T$-dependent parameter. This potential model quantitatively reproduced the lattice simulations on the complex heavy-quark potential in a parameter-independent way provided that screening masses were determined with the background field effective theory. Compared with the hard-thermal-loop perturbation theory, a significant improvement on the nonperturbative properties of the screening mass was achieved based on the effective theory.

Given the complex heavy-quark potential model, in-medium properties of the quarkonium states were obtained by solving the Sch\"ordinger equation, where we focused on the influence of a nontrivial Polyakov loop on the binding and decay of the heavy bound states in a semi-QGP. After taking into account the background field, the binding energies grew dramatically as the temperature approached the deconfining point. This was mainly due to a large increase in the real part of the potential at infinite separation; therefore, the changed amount was roughly the same for all the bound states. On the other hand, although the decay widths showed only a moderate decrease in magnitude, the relative corrections were still significant, indicating a non-negligible effect on quarkonium decay. In general, quarkonium states were more tightly bounded in a partially deconfined medium. However, a notable increase in the dissociation temperature was found only for large-size states, such as $\chi_b(1P)$ and $J/\Psi$, while the dissociation of tightly bounded $\Upsilon(1S)$ was not sensitive to the presence of a background field. In addition, a discontinuity of the order parameter at the deconfining temperature appeared in the effective theory for $SU(3)$ gauge theory. Consequently, we found a jump in both binding energies and decay widths when the phase transition occurred. 

As the background field gave rise to a substantial influence on the binding and decay of quarkonia, it would be interesting to investigate such a ${\cal Q}$ modification on the density evolution of quarkonia during the expansion of the fireball which is expected to give a considerable modification to the final quarkonium spectra. Further work along this line needs to be done in the future.

\section*{Acknowledgments} 

The work is supported by the NSFC of China (Projects No. 12065004 and No. 12465022), and by the Central Government Guidance Funds for Local Scientific and Technological Development, China (No. Guike ZY22096024).

\section*{Data Availability} 

No data were created or analyzed in this study.

\bibliography{paper}

\begin{thebibliography}{30}%
\makeatletter
\providecommand \@ifxundefined [1]{%
 \@ifx{#1\undefined}
}%
\providecommand \@ifnum [1]{%
 \ifnum #1\expandafter \@firstoftwo
 \else \expandafter \@secondoftwo
 \fi
}%
\providecommand \@ifx [1]{%
 \ifx #1\expandafter \@firstoftwo
 \else \expandafter \@secondoftwo
 \fi
}%
\providecommand \natexlab [1]{#1}%
\providecommand \enquote  [1]{``#1''}%
\providecommand \bibnamefont  [1]{#1}%
\providecommand \bibfnamefont [1]{#1}%
\providecommand \citenamefont [1]{#1}%
\providecommand \href@noop [0]{\@secondoftwo}%
\providecommand \href [0]{\begingroup \@sanitize@url \@href}%
\providecommand \@href[1]{\@@startlink{#1}\@@href}%
\providecommand \@@href[1]{\endgroup#1\@@endlink}%
\providecommand \@sanitize@url [0]{\catcode `\\12\catcode `\$12\catcode `\&12\catcode `\#12\catcode `\^12\catcode `\_12\catcode `\%12\relax}%
\providecommand \@@startlink[1]{}%
\providecommand \@@endlink[0]{}%
\providecommand \url  [0]{\begingroup\@sanitize@url \@url }%
\providecommand \@url [1]{\endgroup\@href {#1}{\urlprefix }}%
\providecommand \urlprefix  [0]{URL }%
\providecommand \Eprint [0]{\href }%
\providecommand \doibase [0]{https://doi.org/}%
\providecommand \selectlanguage [0]{\@gobble}%
\providecommand \bibinfo  [0]{\@secondoftwo}%
\providecommand \bibfield  [0]{\@secondoftwo}%
\providecommand \translation [1]{[#1]}%
\providecommand \BibitemOpen [0]{}%
\providecommand \bibitemStop [0]{}%
\providecommand \bibitemNoStop [0]{.\EOS\space}%
\providecommand \EOS [0]{\spacefactor3000\relax}%
\providecommand \BibitemShut  [1]{\csname bibitem#1\endcsname}%
\let\auto@bib@innerbib\@empty
\bibitem [{\citenamefont {Andersen}\ \emph {et~al.}(1999)\citenamefont {Andersen}, \citenamefont {Braaten},\ and\ \citenamefont {Strickland}}]{Andersen:1999fw}%
  \BibitemOpen
  \bibfield  {author} {\bibinfo {author} {\bibfnamefont {J.~O.}\ \bibnamefont {Andersen}}, \bibinfo {author} {\bibfnamefont {E.}~\bibnamefont {Braaten}},\ and\ \bibinfo {author} {\bibfnamefont {M.}~\bibnamefont {Strickland}},\ }\bibfield  {title} {\bibinfo {title} {{Hard thermal loop resummation of the free energy of a hot gluon plasma}},\ }\href {https://doi.org/10.1103/PhysRevLett.83.2139} {\bibfield  {journal} {\bibinfo  {journal} {Phys. Rev. Lett.}\ }\textbf {\bibinfo {volume} {83}},\ \bibinfo {pages} {2139} (\bibinfo {year} {1999})}\BibitemShut {NoStop}%
\bibitem [{\citenamefont {Andersen}\ \emph {et~al.}(2002)\citenamefont {Andersen}, \citenamefont {Braaten}, \citenamefont {Petitgirard},\ and\ \citenamefont {Strickland}}]{Andersen:2002ey}%
  \BibitemOpen
  \bibfield  {author} {\bibinfo {author} {\bibfnamefont {J.~O.}\ \bibnamefont {Andersen}}, \bibinfo {author} {\bibfnamefont {E.}~\bibnamefont {Braaten}}, \bibinfo {author} {\bibfnamefont {E.}~\bibnamefont {Petitgirard}},\ and\ \bibinfo {author} {\bibfnamefont {M.}~\bibnamefont {Strickland}},\ }\bibfield  {title} {\bibinfo {title} {{HTL perturbation theory to two loops}},\ }\href {https://doi.org/10.1103/PhysRevD.66.085016} {\bibfield  {journal} {\bibinfo  {journal} {Phys. Rev. D}\ }\textbf {\bibinfo {volume} {66}},\ \bibinfo {pages} {085016} (\bibinfo {year} {2002})}\BibitemShut {NoStop}%
\bibitem [{\citenamefont {Haque}\ \emph {et~al.}(2008)\citenamefont {Haque}, \citenamefont {Bandyopadhyay}, \citenamefont {Andersen}, \citenamefont {Mustafa}, \citenamefont {Strickland},\ and\ \citenamefont {Su}}]{Haque:2014rua}%
  \BibitemOpen
  \bibfield  {author} {\bibinfo {author} {\bibfnamefont {N.}~\bibnamefont {Haque}}, \bibinfo {author} {\bibfnamefont {A.}~\bibnamefont {Bandyopadhyay}}, \bibinfo {author} {\bibfnamefont {J.~O.}\ \bibnamefont {Andersen}}, \bibinfo {author} {\bibfnamefont {M.~G.}\ \bibnamefont {Mustafa}}, \bibinfo {author} {\bibfnamefont {M.}~\bibnamefont {Strickland}},\ and\ \bibinfo {author} {\bibfnamefont {N.}~\bibnamefont {Su}},\ }\bibfield  {title} {\bibinfo {title} {{Three-loop HTLpt thermodynamics at finite temperature and chemical potential}},\ }\href {https://doi.org/10.1007/JHEP05(2014)027} {\bibfield  {journal} {\bibinfo  {journal} {J. High Energy Phys.}\ }\textbf {\bibinfo {volume} {05}},\ \bibinfo {pages} {(2014) 027}}\BibitemShut {NoStop}%
\bibitem [{\citenamefont {Hidaka}\ and\ \citenamefont {Pisarski}(2008)}]{Hidaka:2008dr}%
  \BibitemOpen
  \bibfield  {author} {\bibinfo {author} {\bibfnamefont {Y.}~\bibnamefont {Hidaka}}\ and\ \bibinfo {author} {\bibfnamefont {R.~D.}\ \bibnamefont {Pisarski}},\ }\bibfield  {title} {\bibinfo {title} {{Suppression of the shear viscosity in a ''semi'' quark gluon plasma}},\ }\href {https://doi.org/10.1103/PhysRevD.78.071501} {\bibfield  {journal} {\bibinfo  {journal} {Phys. Rev. D}\ }\textbf {\bibinfo {volume} {78}},\ \bibinfo {pages} {071501} (\bibinfo {year} {2008})}\BibitemShut {NoStop}%
\bibitem [{\citenamefont {Matsui}\ and\ \citenamefont {Satz}(1986)}]{matsui1986j}%
  \BibitemOpen
  \bibfield  {author} {\bibinfo {author} {\bibfnamefont {T.}~\bibnamefont {Matsui}}\ and\ \bibinfo {author} {\bibfnamefont {H.}~\bibnamefont {Satz}},\ }\bibfield  {title} {\bibinfo {title} {J/$\psi$ suppression by quark-gluon plasma formation},\ }\href@noop {} {\bibfield  {journal} {\bibinfo  {journal} {Phys. Lett. B}\ }\textbf {\bibinfo {volume} {178}},\ \bibinfo {pages} {416} (\bibinfo {year} {1986})}\BibitemShut {NoStop}%
\bibitem [{\citenamefont {Hidaka}\ and\ \citenamefont {Pisarski}(2021)}]{Hidaka:2020vna}%
  \BibitemOpen
  \bibfield  {author} {\bibinfo {author} {\bibfnamefont {Y.}~\bibnamefont {Hidaka}}\ and\ \bibinfo {author} {\bibfnamefont {R.~D.}\ \bibnamefont {Pisarski}},\ }\bibfield  {title} {\bibinfo {title} {{Effective models of a semi-quark-gluon plasma}},\ }\href {https://doi.org/10.1103/PhysRevD.104.074036} {\bibfield  {journal} {\bibinfo  {journal} {Phys. Rev. D}\ }\textbf {\bibinfo {volume} {104}},\ \bibinfo {pages} {074036} (\bibinfo {year} {2021})}\BibitemShut {NoStop}%
\bibitem [{\citenamefont {Dumitru}\ \emph {et~al.}(2011)\citenamefont {Dumitru}, \citenamefont {Guo}, \citenamefont {Hidaka}, \citenamefont {Altes},\ and\ \citenamefont {Pisarski}}]{Dumitru:2010mj}%
  \BibitemOpen
  \bibfield  {author} {\bibinfo {author} {\bibfnamefont {A.}~\bibnamefont {Dumitru}}, \bibinfo {author} {\bibfnamefont {Y.}~\bibnamefont {Guo}}, \bibinfo {author} {\bibfnamefont {Y.}~\bibnamefont {Hidaka}}, \bibinfo {author} {\bibfnamefont {C.~P.~K.}\ \bibnamefont {Altes}},\ and\ \bibinfo {author} {\bibfnamefont {R.~D.}\ \bibnamefont {Pisarski}},\ }\bibfield  {title} {\bibinfo {title} {{How wide is the transition to deconfinement?}},\ }\href {https://doi.org/10.1103/PhysRevD.83.034022} {\bibfield  {journal} {\bibinfo  {journal} {Phys. Rev. D}\ }\textbf {\bibinfo {volume} {83}},\ \bibinfo {pages} {034022} (\bibinfo {year} {2011})}\BibitemShut {NoStop}%
\bibitem [{\citenamefont {Dumitru}\ \emph {et~al.}(2012)\citenamefont {Dumitru}, \citenamefont {Guo}, \citenamefont {Hidaka}, \citenamefont {Altes},\ and\ \citenamefont {Pisarski}}]{Dumitru:2012fw}%
  \BibitemOpen
  \bibfield  {author} {\bibinfo {author} {\bibfnamefont {A.}~\bibnamefont {Dumitru}}, \bibinfo {author} {\bibfnamefont {Y.}~\bibnamefont {Guo}}, \bibinfo {author} {\bibfnamefont {Y.}~\bibnamefont {Hidaka}}, \bibinfo {author} {\bibfnamefont {C.~P.~K.}\ \bibnamefont {Altes}},\ and\ \bibinfo {author} {\bibfnamefont {R.~D.}\ \bibnamefont {Pisarski}},\ }\bibfield  {title} {\bibinfo {title} {{Effective matrix model for deconfinement in pure gauge theories}},\ }\href {https://doi.org/10.1103/PhysRevD.86.105017} {\bibfield  {journal} {\bibinfo  {journal} {Phys. Rev. D}\ }\textbf {\bibinfo {volume} {86}},\ \bibinfo {pages} {105017} (\bibinfo {year} {2012})}\BibitemShut {NoStop}%
\bibitem [{\citenamefont {Guo}(2014)}]{Guo:2014zra}%
  \BibitemOpen
  \bibfield  {author} {\bibinfo {author} {\bibfnamefont {Y.}~\bibnamefont {Guo}},\ }\bibfield  {title} {\bibinfo {title} {{Matrix models for deconfinement and their perturbative corrections}},\ }\href {https://doi.org/10.1007/JHEP11(2014)111} {\bibfield  {journal} {\bibinfo  {journal} {J. High Energy Phys.}\ }\textbf {\bibinfo {volume} {11}},\ \bibinfo {pages} {(2014) 111}}\BibitemShut {NoStop}%
\bibitem [{\citenamefont {Brambilla}\ \emph {et~al.}(2000)\citenamefont {Brambilla}, \citenamefont {Pineda}, \citenamefont {Soto},\ and\ \citenamefont {Vairo}}]{Brambilla:1999xf}%
  \BibitemOpen
  \bibfield  {author} {\bibinfo {author} {\bibfnamefont {N.}~\bibnamefont {Brambilla}}, \bibinfo {author} {\bibfnamefont {A.}~\bibnamefont {Pineda}}, \bibinfo {author} {\bibfnamefont {J.}~\bibnamefont {Soto}},\ and\ \bibinfo {author} {\bibfnamefont {A.}~\bibnamefont {Vairo}},\ }\bibfield  {title} {\bibinfo {title} {{Potential NRQCD: An Effective theory for heavy quarkonium}},\ }\href {https://doi.org/10.1016/S0550-3213(99)00693-8} {\bibfield  {journal} {\bibinfo  {journal} {Nucl. Phys.}\ }\textbf {\bibinfo {volume} {B566}},\ \bibinfo {pages} {275} (\bibinfo {year} {2000})}\BibitemShut {NoStop}%
\bibitem [{\citenamefont {Brambilla}\ \emph {et~al.}(2005)\citenamefont {Brambilla}, \citenamefont {Pineda}, \citenamefont {Soto},\ and\ \citenamefont {Vairo}}]{Brambilla:2004jw}%
  \BibitemOpen
  \bibfield  {author} {\bibinfo {author} {\bibfnamefont {N.}~\bibnamefont {Brambilla}}, \bibinfo {author} {\bibfnamefont {A.}~\bibnamefont {Pineda}}, \bibinfo {author} {\bibfnamefont {J.}~\bibnamefont {Soto}},\ and\ \bibinfo {author} {\bibfnamefont {A.}~\bibnamefont {Vairo}},\ }\bibfield  {title} {\bibinfo {title} {{Effective field theories for heavy quarkonium}},\ }\href {https://doi.org/10.1103/RevModPhys.77.1423} {\bibfield  {journal} {\bibinfo  {journal} {Rev. Mod. Phys.}\ }\textbf {\bibinfo {volume} {77}},\ \bibinfo {pages} {1423} (\bibinfo {year} {2005})}\BibitemShut {NoStop}%
\bibitem [{\citenamefont {Burnier}\ and\ \citenamefont {Rothkopf}(2017)}]{Burnier:2016mxc}%
  \BibitemOpen
  \bibfield  {author} {\bibinfo {author} {\bibfnamefont {Y.}~\bibnamefont {Burnier}}\ and\ \bibinfo {author} {\bibfnamefont {A.}~\bibnamefont {Rothkopf}},\ }\bibfield  {title} {\bibinfo {title} {{Complex heavy-quark potential and Debye mass in a gluonic medium from lattice QCD}},\ }\href {https://doi.org/10.1103/PhysRevD.95.054511} {\bibfield  {journal} {\bibinfo  {journal} {Phys. Rev. D}\ }\textbf {\bibinfo {volume} {95}},\ \bibinfo {pages} {054511} (\bibinfo {year} {2017})}\BibitemShut {NoStop}%
\bibitem [{\citenamefont {Bazavov}\ \emph {et~al.}(2024)\citenamefont {Bazavov}, \citenamefont {Hoying}, \citenamefont {Larsen}, \citenamefont {Mukherjee}, \citenamefont {Petreczky}, \citenamefont {Rothkopf},\ and\ \citenamefont {Weber}}]{Bazavov:2023dci}%
  \BibitemOpen
  \bibfield  {author} {\bibinfo {author} {\bibfnamefont {A.}~\bibnamefont {Bazavov}}, \bibinfo {author} {\bibfnamefont {D.}~\bibnamefont {Hoying}}, \bibinfo {author} {\bibfnamefont {R.~N.}\ \bibnamefont {Larsen}}, \bibinfo {author} {\bibfnamefont {S.}~\bibnamefont {Mukherjee}}, \bibinfo {author} {\bibfnamefont {P.}~\bibnamefont {Petreczky}}, \bibinfo {author} {\bibfnamefont {A.}~\bibnamefont {Rothkopf}},\ and\ \bibinfo {author} {\bibfnamefont {J.~H.}\ \bibnamefont {Weber}} (\bibinfo {collaboration} {HotQCD Collaboration}),\ }\bibfield  {title} {\bibinfo {title} {{Unscreened forces in the quark-gluon plasma?}},\ }\href {https://doi.org/10.1103/PhysRevD.109.074504} {\bibfield  {journal} {\bibinfo  {journal} {Phys. Rev. D}\ }\textbf {\bibinfo {volume} {109}},\ \bibinfo {pages} {074504} (\bibinfo {year} {2024})}\BibitemShut {NoStop}%
\bibitem [{\citenamefont {Larsen}\ \emph {et~al.}(2024)\citenamefont {Larsen}, \citenamefont {Parkar}, \citenamefont {Rothkopf},\ and\ \citenamefont {Weber}}]{Larsen:2024wgw}%
  \BibitemOpen
  \bibfield  {author} {\bibinfo {author} {\bibfnamefont {R.~N.}\ \bibnamefont {Larsen}}, \bibinfo {author} {\bibfnamefont {G.}~\bibnamefont {Parkar}}, \bibinfo {author} {\bibfnamefont {A.}~\bibnamefont {Rothkopf}},\ and\ \bibinfo {author} {\bibfnamefont {J.~H.}\ \bibnamefont {Weber}},\ }\bibfield  {title} {\bibinfo {title} {{In-medium static inter-quark potential on high resolution quenched lattices}},\ }\href {https://doi.org/10.1103/PhysRevD.110.114501} {\bibfield  {journal} {\bibinfo  {journal} {Phys. Rev. D}\ }\textbf {\bibinfo {volume} {110}},\ \bibinfo {pages} {114501} (\bibinfo {year} {2024})}\BibitemShut {NoStop}%
\bibitem [{\citenamefont {Shi}\ \emph {et~al.}(2022)\citenamefont {Shi}, \citenamefont {Zhou}, \citenamefont {Zhao}, \citenamefont {Mukherjee},\ and\ \citenamefont {Zhuang}}]{Shi:2021qri}%
  \BibitemOpen
  \bibfield  {author} {\bibinfo {author} {\bibfnamefont {S.}~\bibnamefont {Shi}}, \bibinfo {author} {\bibfnamefont {K.}~\bibnamefont {Zhou}}, \bibinfo {author} {\bibfnamefont {J.}~\bibnamefont {Zhao}}, \bibinfo {author} {\bibfnamefont {S.}~\bibnamefont {Mukherjee}},\ and\ \bibinfo {author} {\bibfnamefont {P.}~\bibnamefont {Zhuang}},\ }\bibfield  {title} {\bibinfo {title} {{Heavy quark potential in the quark-gluon plasma: Deep neural network meets lattice quantum chromodynamics}},\ }\href {https://doi.org/10.1103/PhysRevD.105.014017} {\bibfield  {journal} {\bibinfo  {journal} {Phys. Rev. D}\ }\textbf {\bibinfo {volume} {105}},\ \bibinfo {pages} {014017} (\bibinfo {year} {2022})}\BibitemShut {NoStop}%
\bibitem [{\citenamefont {Bala}\ \emph {et~al.}(2022)\citenamefont {Bala}, \citenamefont {Kaczmarek}, \citenamefont {Larsen}, \citenamefont {Mukherjee}, \citenamefont {Parkar}, \citenamefont {Petreczky}, \citenamefont {Rothkopf},\ and\ \citenamefont {Weber}}]{Bala:2021fkm}%
  \BibitemOpen
  \bibfield  {author} {\bibinfo {author} {\bibfnamefont {D.}~\bibnamefont {Bala}}, \bibinfo {author} {\bibfnamefont {O.}~\bibnamefont {Kaczmarek}}, \bibinfo {author} {\bibfnamefont {R.}~\bibnamefont {Larsen}}, \bibinfo {author} {\bibfnamefont {S.}~\bibnamefont {Mukherjee}}, \bibinfo {author} {\bibfnamefont {G.}~\bibnamefont {Parkar}}, \bibinfo {author} {\bibfnamefont {P.}~\bibnamefont {Petreczky}}, \bibinfo {author} {\bibfnamefont {A.}~\bibnamefont {Rothkopf}},\ and\ \bibinfo {author} {\bibfnamefont {J.~H.}\ \bibnamefont {Weber}} (\bibinfo {collaboration} {HotQCD Collaboration}),\ }\bibfield  {title} {\bibinfo {title} {{Static quark-antiquark interactions at nonzero temperature from lattice QCD}},\ }\href {https://doi.org/10.1103/PhysRevD.105.054513} {\bibfield  {journal} {\bibinfo  {journal} {Phys. Rev. D}\ }\textbf {\bibinfo {volume} {105}},\ \bibinfo {pages} {054513} (\bibinfo {year} {2022})}\BibitemShut {NoStop}%
\bibitem [{\citenamefont {Karsch}\ \emph {et~al.}(1988)\citenamefont {Karsch}, \citenamefont {Mehr},\ and\ \citenamefont {Satz}}]{Karsch:1987pv}%
  \BibitemOpen
  \bibfield  {author} {\bibinfo {author} {\bibfnamefont {F.}~\bibnamefont {Karsch}}, \bibinfo {author} {\bibfnamefont {M.~T.}\ \bibnamefont {Mehr}},\ and\ \bibinfo {author} {\bibfnamefont {H.}~\bibnamefont {Satz}},\ }\bibfield  {title} {\bibinfo {title} {{Color screening and deconfinement for bound states of heavy quarks}},\ }\href {https://doi.org/10.1007/BF01549722} {\bibfield  {journal} {\bibinfo  {journal} {Z. Phys. C}\ }\textbf {\bibinfo {volume} {37}},\ \bibinfo {pages} {617} (\bibinfo {year} {1988})}\BibitemShut {NoStop}%
\bibitem [{\citenamefont {Guo}\ \emph {et~al.}(2019)\citenamefont {Guo}, \citenamefont {Dong}, \citenamefont {Pan},\ and\ \citenamefont {Moldes}}]{Guo:2018vwy}%
  \BibitemOpen
  \bibfield  {author} {\bibinfo {author} {\bibfnamefont {Y.}~\bibnamefont {Guo}}, \bibinfo {author} {\bibfnamefont {L.}~\bibnamefont {Dong}}, \bibinfo {author} {\bibfnamefont {J.}~\bibnamefont {Pan}},\ and\ \bibinfo {author} {\bibfnamefont {M.~R.}\ \bibnamefont {Moldes}},\ }\bibfield  {title} {\bibinfo {title} {{Modelling the non-perturbative contributions to the complex heavy-quark potential}},\ }\href {https://doi.org/10.1103/PhysRevD.100.036011} {\bibfield  {journal} {\bibinfo  {journal} {Phys. Rev. D}\ }\textbf {\bibinfo {volume} {100}},\ \bibinfo {pages} {036011} (\bibinfo {year} {2019})}\BibitemShut {NoStop}%
\bibitem [{\citenamefont {Chetyrkin}\ \emph {et~al.}(1999)\citenamefont {Chetyrkin}, \citenamefont {Narison},\ and\ \citenamefont {Zakharov}}]{Chetyrkin:1998yr}%
  \BibitemOpen
  \bibfield  {author} {\bibinfo {author} {\bibfnamefont {K.~G.}\ \bibnamefont {Chetyrkin}}, \bibinfo {author} {\bibfnamefont {S.}~\bibnamefont {Narison}},\ and\ \bibinfo {author} {\bibfnamefont {V.~I.}\ \bibnamefont {Zakharov}},\ }\bibfield  {title} {\bibinfo {title} {{Short distance tachyonic gluon mass and 1/$Q^2$ corrections}},\ }\href {https://doi.org/10.1016/S0550-3213(99)00167-4} {\bibfield  {journal} {\bibinfo  {journal} {Nucl. Phys.}\ }\textbf {\bibinfo {volume} {B550}},\ \bibinfo {pages} {353} (\bibinfo {year} {1999})}\BibitemShut {NoStop}%
\bibitem [{\citenamefont {Meisinger}\ \emph {et~al.}(2002)\citenamefont {Meisinger}, \citenamefont {Miller},\ and\ \citenamefont {Ogilvie}}]{Meisinger:2001cq}%
  \BibitemOpen
  \bibfield  {author} {\bibinfo {author} {\bibfnamefont {P.~N.}\ \bibnamefont {Meisinger}}, \bibinfo {author} {\bibfnamefont {T.~R.}\ \bibnamefont {Miller}},\ and\ \bibinfo {author} {\bibfnamefont {M.~C.}\ \bibnamefont {Ogilvie}},\ }\bibfield  {title} {\bibinfo {title} {{Phenomenological equations of state for the quark gluon plasma}},\ }\href {https://doi.org/10.1103/PhysRevD.65.034009} {\bibfield  {journal} {\bibinfo  {journal} {Phys. Rev. D}\ }\textbf {\bibinfo {volume} {65}},\ \bibinfo {pages} {034009} (\bibinfo {year} {2002})}\BibitemShut {NoStop}%
\bibitem [{\citenamefont {Guo}\ and\ \citenamefont {Kuang}(2021)}]{Guo:2020jvc}%
  \BibitemOpen
  \bibfield  {author} {\bibinfo {author} {\bibfnamefont {Y.}~\bibnamefont {Guo}}\ and\ \bibinfo {author} {\bibfnamefont {Z.}~\bibnamefont {Kuang}},\ }\bibfield  {title} {\bibinfo {title} {{Resummed gluon propagator and Debye screening effect in a holonomous plasma}},\ }\href {https://doi.org/10.1103/PhysRevD.104.014015} {\bibfield  {journal} {\bibinfo  {journal} {Phys. Rev. D}\ }\textbf {\bibinfo {volume} {104}},\ \bibinfo {pages} {014015} (\bibinfo {year} {2021})}\BibitemShut {NoStop}%
\bibitem [{\citenamefont {Hidaka}\ and\ \citenamefont {Pisarski}(2009)}]{Hidaka:2009hs}%
  \BibitemOpen
  \bibfield  {author} {\bibinfo {author} {\bibfnamefont {Y.}~\bibnamefont {Hidaka}}\ and\ \bibinfo {author} {\bibfnamefont {R.~D.}\ \bibnamefont {Pisarski}},\ }\bibfield  {title} {\bibinfo {title} {{Hard thermal loops, to quadratic order, in the background of a spatial 't Hooft loop}},\ }\href {https://doi.org/10.1103/PhysRevD.80.036004} {\bibfield  {journal} {\bibinfo  {journal} {Phys. Rev. D}\ }\textbf {\bibinfo {volume} {80}},\ \bibinfo {pages} {036004} (\bibinfo {year} {2009})},\ \bibinfo {note} {[Erratum: Phys.Rev.D 102, 059902 (2020)]}\BibitemShut {NoStop}%
\bibitem [{\citenamefont {Guo}\ and\ \citenamefont {Du}()}]{Guo:2018scp}%
  \BibitemOpen
  \bibfield  {author} {\bibinfo {author} {\bibfnamefont {Y.}~\bibnamefont {Guo}}\ and\ \bibinfo {author} {\bibfnamefont {Q.}~\bibnamefont {Du}},\ }\bibfield  {title} {\bibinfo {title} {{Two-loop perturbative corrections to the constrained effective potential in thermal QCD}},\ }\href {https://doi.org/10.1007/JHEP05(2019)042} {\bibfield  {journal} {\bibinfo  {journal} {J. High Energy Phys.}\ }\textbf {\bibinfo {volume} {05}},\ \bibinfo {pages} {(2019) 042}}\BibitemShut {NoStop}%
\bibitem [{\citenamefont {Wang}\ \emph {et~al.}(2022)\citenamefont {Wang}, \citenamefont {Du},\ and\ \citenamefont {Guo}}]{Wang:2022dcw}%
  \BibitemOpen
  \bibfield  {author} {\bibinfo {author} {\bibfnamefont {Y.}~\bibnamefont {Wang}}, \bibinfo {author} {\bibfnamefont {Q.}~\bibnamefont {Du}},\ and\ \bibinfo {author} {\bibfnamefont {Y.}~\bibnamefont {Guo}},\ }\bibfield  {title} {\bibinfo {title} {{Real-time hard-thermal-loop gluon self-energy in a semiquark-gluon plasma}},\ }\href {https://doi.org/10.1103/PhysRevD.106.054033} {\bibfield  {journal} {\bibinfo  {journal} {Phys. Rev. D}\ }\textbf {\bibinfo {volume} {106}},\ \bibinfo {pages} {054033} (\bibinfo {year} {2022})}\BibitemShut {NoStop}%
\bibitem [{\citenamefont {Dumitru}\ \emph {et~al.}(2009)\citenamefont {Dumitru}, \citenamefont {Guo}, \citenamefont {Mocsy},\ and\ \citenamefont {Strickland}}]{Dumitru:2009ni}%
  \BibitemOpen
  \bibfield  {author} {\bibinfo {author} {\bibfnamefont {A.}~\bibnamefont {Dumitru}}, \bibinfo {author} {\bibfnamefont {Y.}~\bibnamefont {Guo}}, \bibinfo {author} {\bibfnamefont {A.}~\bibnamefont {Mocsy}},\ and\ \bibinfo {author} {\bibfnamefont {M.}~\bibnamefont {Strickland}},\ }\bibfield  {title} {\bibinfo {title} {{Quarkonium states in an anisotropic QCD plasma}},\ }\href {https://doi.org/10.1103/PhysRevD.79.054019} {\bibfield  {journal} {\bibinfo  {journal} {Phys. Rev. D}\ }\textbf {\bibinfo {volume} {79}},\ \bibinfo {pages} {054019} (\bibinfo {year} {2009})}\BibitemShut {NoStop}%
\bibitem [{\citenamefont {Du}\ \emph {et~al.}(2024)\citenamefont {Du}, \citenamefont {Du},\ and\ \citenamefont {Guo}}]{Du:2024riq}%
  \BibitemOpen
  \bibfield  {author} {\bibinfo {author} {\bibfnamefont {Q.}~\bibnamefont {Du}}, \bibinfo {author} {\bibfnamefont {M.}~\bibnamefont {Du}},\ and\ \bibinfo {author} {\bibfnamefont {Y.}~\bibnamefont {Guo}},\ }\bibfield  {title} {\bibinfo {title} {{Collisional energy loss of a heavy quark in a semiquark-gluon plasma}},\ }\href {https://doi.org/10.1103/PhysRevD.110.034011} {\bibfield  {journal} {\bibinfo  {journal} {Phys. Rev. D}\ }\textbf {\bibinfo {volume} {110}},\ \bibinfo {pages} {034011} (\bibinfo {year} {2024})}\BibitemShut {NoStop}%
\bibitem [{\citenamefont {Thakur}\ \emph {et~al.}(2014)\citenamefont {Thakur}, \citenamefont {Kakade},\ and\ \citenamefont {Patra}}]{Thakur:2013nia}%
  \BibitemOpen
  \bibfield  {author} {\bibinfo {author} {\bibfnamefont {L.}~\bibnamefont {Thakur}}, \bibinfo {author} {\bibfnamefont {U.}~\bibnamefont {Kakade}},\ and\ \bibinfo {author} {\bibfnamefont {B.~K.}\ \bibnamefont {Patra}},\ }\bibfield  {title} {\bibinfo {title} {{Dissociation of quarkonium in a complex potential}},\ }\href {https://doi.org/10.1103/PhysRevD.89.094020} {\bibfield  {journal} {\bibinfo  {journal} {Phys. Rev. D}\ }\textbf {\bibinfo {volume} {89}},\ \bibinfo {pages} {094020} (\bibinfo {year} {2014})}\BibitemShut {NoStop}%
\bibitem [{\citenamefont {Burnier}\ and\ \citenamefont {Rothkopf}(2016)}]{Burnier:2015nsa}%
  \BibitemOpen
  \bibfield  {author} {\bibinfo {author} {\bibfnamefont {Y.}~\bibnamefont {Burnier}}\ and\ \bibinfo {author} {\bibfnamefont {A.}~\bibnamefont {Rothkopf}},\ }\bibfield  {title} {\bibinfo {title} {{A gauge invariant Debye mass and the complex heavy-quark potential}},\ }\href {https://doi.org/10.1016/j.physletb.2015.12.031} {\bibfield  {journal} {\bibinfo  {journal} {Phys. Lett. B}\ }\textbf {\bibinfo {volume} {753}},\ \bibinfo {pages} {232} (\bibinfo {year} {2016})}\BibitemShut {NoStop}%
\bibitem [{\citenamefont {Strickland}\ and\ \citenamefont {Yager-Elorriaga}(2010)}]{Strickland:2009ft}%
  \BibitemOpen
  \bibfield  {author} {\bibinfo {author} {\bibfnamefont {M.}~\bibnamefont {Strickland}}\ and\ \bibinfo {author} {\bibfnamefont {D.}~\bibnamefont {Yager-Elorriaga}},\ }\bibfield  {title} {\bibinfo {title} {{A parallel algorithm for solving the 3d Schrodinger equation}},\ }\href {https://doi.org/10.1016/j.jcp.2010.04.032} {\bibfield  {journal} {\bibinfo  {journal} {J. Comput. Phys.}\ }\textbf {\bibinfo {volume} {229}},\ \bibinfo {pages} {6015} (\bibinfo {year} {2010})}\BibitemShut {NoStop}%
\bibitem [{\citenamefont {Delgado}\ \emph {et~al.}(2022)\citenamefont {Delgado}, \citenamefont {Steinbei\ss{}er}, \citenamefont {Strickland},\ and\ \citenamefont {Weber}}]{Delgado:2020ozh}%
  \BibitemOpen
  \bibfield  {author} {\bibinfo {author} {\bibfnamefont {R.~L.}\ \bibnamefont {Delgado}}, \bibinfo {author} {\bibfnamefont {S.}~\bibnamefont {Steinbei\ss{}er}}, \bibinfo {author} {\bibfnamefont {M.}~\bibnamefont {Strickland}},\ and\ \bibinfo {author} {\bibfnamefont {J.~H.}\ \bibnamefont {Weber}},\ }\bibfield  {title} {\bibinfo {title} {{The relativistic Schr\"odinger equation through FFTW 3: An extension of quantumfdtd}},\ }\href {https://doi.org/10.1016/j.cpc.2021.108250} {\bibfield  {journal} {\bibinfo  {journal} {Comput. Phys. Commun.}\ }\textbf {\bibinfo {volume} {272}},\ \bibinfo {pages} {108250} (\bibinfo {year} {2022})}\BibitemShut {NoStop}%
\end{thebibliography}%

\end{document}